\documentclass{article}

\newif\iftwocolumnlayout
\twocolumnlayouttrue

\newif\ifcombinedsupplement
\combinedsupplementtrue

\newcommand{\uniformtablesetup}{%
  \small
  \setlength{\tabcolsep}{3.5pt}%
  \renewcommand{\arraystretch}{1.12}%
}

\usepackage[british]{babel}
\usepackage{authblk}
\usepackage[utf8]{inputenc}
\usepackage{subcaption}

\iftwocolumnlayout
  
  \usepackage[letterpaper,top=1.65cm,bottom=1.75cm,left=1.55cm,right=1.55cm]{geometry}
  \usepackage{dblfloatfix}
  \usepackage{balance}
  \usepackage{multicol}
  \usepackage{cuted}
  \usepackage{titlesec}
  \titlespacing*{\subsection}{0pt}{2.3ex}{0.8ex}
  \titlespacing*{\subsubsection}{0pt}{2.0ex}{0.6ex}
  \newcommand{\prepareappendix}{\clearpage\onecolumn}
  \newcommand{\startreferencecolumns}{\begin{multicols}{2}}
  \newcommand{\stopreferencecolumns}{\end{multicols}}
\else
  
  \usepackage[letterpaper,top=2cm,bottom=2cm,left=3cm,right=3cm]{geometry}
  \newcommand{\prepareappendix}{\FloatBarrier}
  \newcommand{\startreferencecolumns}{}
  \newcommand{\stopreferencecolumns}{}
\fi

\usepackage{amsmath}
\usepackage{microtype}
\AtBeginDocument{\setlength{\parfillskip}{0pt plus 0.8\columnwidth}}
\usepackage{flafter}
\usepackage{float}
\usepackage{placeins}
\usepackage{comment}
\usepackage{graphicx}
\usepackage{adjustbox}
\usepackage[colorlinks=true, allcolors=blue]{hyperref}
\usepackage{xr-hyper}

\usepackage{multirow}
\usepackage{array}
\usepackage{booktabs}
\usepackage{tabularx}
\usepackage{makecell,ragged2e}
\usepackage{amssymb}
\usepackage[font={small,color=black!80}]{caption}
\usepackage{xcolor}

\definecolor{darkred}{RGB}{150,0,0}

\iftwocolumnlayout
  \newenvironment{manuscriptfigure}{\begin{figure*}[!tb]}{\end{figure*}}
  \newenvironment{manuscripttable}{\begin{table*}[!tb]}{\end{table*}}
\else
  \newenvironment{manuscriptfigure}{\begin{figure}[!tbp]}{\end{figure}}
  \newenvironment{manuscripttable}{\begin{table}[!tbp]}{\end{table}}
\fi

\iftwocolumnlayout
  \newenvironment{pipelinefigure}{\begin{figure*}[!t]\centering}{\end{figure*}}
  \newcommand{\pipelinecaption}[1]{\caption{#1}}
\else
  \newenvironment{pipelinefigure}{\begin{figure}[!htbp]\centering}{\end{figure}}
  \newcommand{\pipelinecaption}[1]{\caption{#1}}
\fi
\iftwocolumnlayout
  \newenvironment{appendixfigure}{\begin{figure}[!htbp]}{\end{figure}}
  \newenvironment{appendixtable}{\begin{table}[!htbp]}{\end{table}}
  \newenvironment{appendixassetfigure}{\begin{figure}[!htbp]}{\end{figure}}
  \newenvironment{appendixassettable}{\begin{table}[!htbp]}{\end{table}}
  \newenvironment{appendixlatefigure}{\begin{figure}[!htbp]}{\end{figure}}

\else
  \newenvironment{appendixfigure}{\begin{figure}[!htbp]}{\end{figure}}
  
  \newenvironment{appendixassetfigure}{\begin{figure}[!htbp]}{\end{figure}}
  
  \newenvironment{appendixlatefigure}{\begin{figure}[!htbp]}{\end{figure}}

\fi
\usepackage{acronym}

\newacro{PD}{Parkinson's disease}
\newacro{SPEM}{Smooth Pursuit Eye Movements}
\newacro{SNr}{Substantia nigra pars reticulata}
\newacro{EEG}{Electroencephalogram}
\newacro{ECG}{Electrocardiogram}
\newacro{LLE}{Largest Lyapunov Exponent}
\newacro{CD}{Correlation Dimension}
\newacro{FD}{Fractal Dimension}
\newacro{SVM}{Support Vector Machine}
\newacro{LZC}{Lempel--Ziv Complexity}
\newacro{HC}{Healthy Control}
\newacro{YC}{Young Control}
\newacro{UPDRS}{Unified Parkinson's Disease Rating Scale}
\newacro{MoCA}{Montreal Cognitive Assessment}
\newacro{SPT}{Smooth Pursuit Task}
\newacro{PI}{Permutation Importance}
\newacro{CERM}{Cyclopean Eye Residual Movement}
\newacro{BW}{Baseline Wander}
\newacro{LOPO}{Leave-One-Participant-Out}
\newacro{EER}{Equal Error Rate}
\newacro{SSA}{Singular Spectrum Analysis}
\newacro{EWT}{Empirical Wavelet Transform}
\newacro{ICA}{Independent Component Analysis}
\newacro{ApEn}{Approximate Entropy}
\newacro{cApEn}{Corrected Approximate Entropy}
\newacro{SampEn}{Sample Entropy}
\newacro{QSE}{Quadratic Sample Entropy}
\newacro{ShEn}{Shannon Entropy}
\newacro{MaxEn}{Maximum Entropy}
\newacro{DiffEn}{Differential Entropy}
\newacro{PowEn}{Power Entropy}
\newacro{TE}{Tsallis Entropy}
\newacro{HEn}{Hartley Entropy}
\newacro{CEn}{Collision Entropy}
\newacro{PE}{Permutation Entropy}
\newacro{wPE}{Weighted Permutation Entropy}
\newacro{cPE}{Conditional Permutation Entropy}
\newacro{cwPE}{Conditional Weighted Permutation Entropy}
\newacro{cRPE}{Conditional Rényi Permutation Entropy}
\newacro{FGPE}{Fine-grained Permutation Entropy}
\newacro{mPE}{Modified Permutation Entropy}
\newacro{AAPE}{Amplitude-aware Permutation Entropy}
\newacro{ePE}{Edge Permutation Entropy}
\newacro{UQPE}{Uniform Quantisation-based Permutation Entropy}
\newacro{KFD}{Katz's Fractal Dimension}
\newacro{PFD}{Petrosian Fractal Dimension}
\newacro{DFA}{Detrended Fluctuation Analysis}
\newacro{HE}{Hurst Exponent}
\newacro{AMI}{Average Mutual Information}
\newacro{FNN}{False Nearest Neighbours}
\newacro{ROC}{Receiver Operating Characteristic}
\newacro{AUC}{Area Under the Curve}
\newacro{SGKF}{Stratified Group $k$-Fold}
\newacro{MoE}{Mixture of Experts}
\newacro{CV}{Cross-Validation}
\newacro{RBF}{Radial Basis Function}
\newacro{FFS}{Forward Feature Selection}
\newacro{HFS}{Hierarchical Feature Selection}
\newacro{EDel}{Embedding Delay}
\newacro{EDim}{Embedding Dimension}
\newacro{EF}{Experimental Framework}
\newacro{DS}{Delay Strategy}
\newacroplural{DS}{Delay Strategies}
\newacro{RFS}{Representative Feature Set}

\title{Automated screening of Parkinson's Disease using Smooth Pursuit Eye Movements}
\date{}

\author[1]{J. C. Puerta-Acevedo}
\author[1]{M. F. Alcala-Durand}
\author[1]{J. D. Arias-Londoño}
\author[2]{E. Luque-Buzo}
\author[2]{F. Grandas-Pérez}
\author[1]{J. I. Godino-Llorente\thanks{Corresponding author: J. I. Godino-Llorente (e-mail: \texttt{ignacio.godino@upm.es}). All authors contributed in the same manner.}}
\affil[1]{ETSI Telecomunicación. Universidad Politécnica de Madrid. Spain}
\affil[2]{Movement Disorders Unit, Neurology Department, Hospital General Universitario Gregorio Marañón, Madrid. Spain}

\begin{document}

\long\def\manuscriptabstract{%
Smooth pursuit eye movements are frequently altered in Parkinson's disease, but previous studies have relied mainly on conventional parameters that describe the relationship between gaze and target. This study investigates whether complexity and nonlinear dynamics measures of the gaze position sequences can support the automatic screening of Parkinson's disease. Gaze position sequences were recorded with a video eye tracker in 41 participants with early Parkinson's disease and 47 healthy controls during 12 smooth pursuit tasks. Four experimental frameworks and four embedding delay strategies were compared. The selected experimental setting combined baseline wander removal with a fixed embedding delay. After feature selection, a support vector machine was trained for each task and axis and their outputs were combined in a mixture of experts. The pipeline was developed in 70 participants and evaluated on an external test set of 18 participants. On the external test set, the pipeline obtained an area under the receiver operating characteristic curve of $0.86$. The most informative features described less persistent and more irregular gaze position sequences in Parkinson's disease, mainly along the horizontal axis. These results support complexity and nonlinear dynamics measures of smooth pursuit as candidate markers for Parkinson's disease screening, pending validation in larger independent cohorts.
}%

\iftwocolumnlayout
\makeatletter
\twocolumn[
\begin{@twocolumnfalse}
\maketitle
\begin{abstract}
\manuscriptabstract
\end{abstract}
\vspace{0.8em}
\end{@twocolumnfalse}
]
\makeatother
\else
\maketitle
\begin{abstract}
\manuscriptabstract
\end{abstract}
\fi

\section*{Introduction}

\ac{PD} is a progressive multisystem neurodegenerative disorder characterised mainly by degeneration of dopaminergic neurons in the substantia nigra pars compacta, although several nondopaminergic systems are also affected. It produces motor and nonmotor manifestations, some of which may precede the cardinal motor signs. Among fine motor symptoms, oculomotor alterations are known to occur from the early stages of the disease. In this regard, two-thirds of patients with \ac{PD} showed oculomotor alterations in an electro-oculographic study~\cite{cipparrone1988electro}.

Since eye movements can be recorded quantitatively and noninvasively, they have been investigated as complementary markers of \ac{PD}~\cite{Antoniades2024EyeMovements}. It is known that \ac{PD} introduces certain deviations in the behaviour of the oculomotor system, affecting fixation, saccades, vergence and \ac{SPEM}~\cite{Antoniades2024EyeMovements,frei2021abnormalities}. 

\ac{SPEM} are continuous eye movements that keep the image of a moving target close to the fovea. Their control involves sensory processing, attention, prediction and executive mechanisms, with coordinated activity across cortical, cerebellar, basal ganglia and brainstem circuits~\cite{fukushima2017clinical}. Pursuit-related activity has also been reported in the substantia nigra pars reticulata, a major output nucleus of the basal ganglia~\cite{basso2005activity}. The involvement of the basal ganglia provides a neurophysiological rationale for investigating \ac{SPEM} alterations in \ac{PD}~\cite{frei2021abnormalities}. 

Consistent with this, abnormal saccades during \ac{SPEM} have been reported in $43.8\%$ of patients with \ac{PD} compared with $10.0\%$ of \acp{HC}~\cite{zhou2021differential}. Therefore, \ac{SPEM} analysis is considered of clinical interest for evaluating patients with \ac{PD}~\cite{frei2021abnormalities}.

Deviations in the eye's positions during \ac{SPEM} are typically corrected by means of saccadic movements. Saccades are rapid eye movements that abruptly redirect the gaze to an object of interest. Although the presence of saccadic movements in \ac{SPEM} is considered normal, the frequency, extent and type are different in patients with \ac{PD}~\cite{zhou2021differential}. During \ac{SPEM}, several types of abnormal and involuntary saccadic movements exist in the recorded sequences: catch-up saccades, back-up saccades and square-wave jerks.

Clinical studies in \ac{PD} have extensively investigated a set of conventional and physiologically interpretable parameters extracted from \ac{SPEM} recordings. Pursuit gain, defined as the ratio of eye velocity to target velocity, is the most widely studied~\cite{frei2021abnormalities}. Other parameters include pursuit latency, phase lag, position error, eye velocity, range of motion and the number or amplitude of saccades during pursuit~\cite{rottach1996dynamic,wu2018eye,Popovic2026SmoothPursuitMemorySaccades}.

Saccadic activity during pursuit can also describe the tracking behaviour. Corrective saccades reduce the error between gaze and target position. Additionally, catch-up saccades, for example, act when gaze falls behind the target~\cite{DeBrouwer2002CatchUpSaccades}. On the other hand, anticipatory saccades, back-up saccades and other saccadic intrusions arise from different mechanisms and may differ in number, amplitude, timing and direction between participants with \ac{PD} and \acp{HC}~\cite{zhou2021differential}. %

Reduced gain and saccadic or fragmented pursuit are also among the most frequently reported impairments in \ac{PD}, but findings vary with the stimulus, recording system, medication state and analysis procedure~\cite{frei2021abnormalities}. Newly diagnosed and untreated participants with \ac{PD} have shown alterations associated in some cases with disease duration and motor severity~\cite{Zhou2022DeNovo}, whereas pursuit can remain relatively preserved under specific task conditions~\cite{Fooken2022Preservation}. Recent studies have also reported effects of movement direction, target frequency and medication state, together with alterations in velocity, phase and range of motion in early \ac{PD}~\cite{Zhao2025SPEMLevodopa,Popovic2026SmoothPursuitMemorySaccades}. Pursuit impairment therefore cannot be characterised by gain alone and may depend on the direction, frequency, amplitude, predictability and duration of the target trajectory.

Although the aforementioned characteristics have provided substantial evidence of altered tracking in \ac{PD}, they mainly describe very specific aspects of the relationship between gaze and the target, not taking the temporal organisation and dynamics of the signal into account. On the other hand, a common assumption in biosignal processing is that the presence of a disease can alter the organisation and adaptability of biological dynamics. %
Thus, the associated changes may instead appear as deviations in the regularity, altered persistence or different temporal organisation~\cite{goldberger2002physiologic}.

This motivates the investigation of additional measures capable of describing the organisation and dynamics of the gaze position sequence itself using complexity and nonlinear dynamics measures. In this regard, complexity and nonlinear analyses are common methods used for characterisation and parametrisation of different biomedical signals derived from nonlinear dynamic systems. Complex, irregular and nonstationary dynamical systems represent many biological objects and physiological events, but detecting their relevant hidden properties is not straightforward because no clear rules describe their behaviour. Thus, exploring the observations of these systems requires appropriate and specific methods. In this context, nonlinear time series analysis is often used as an indirect approach to study the system's structure and dynamics.

The oculomotor system, which controls the muscles that move the eyes, is also a complex nonlinear dynamical system. Under the assumption that the monitoring of the oculomotor system can be carried out indirectly using the gaze positions obtained by an eye tracker device, the underlying hypothesis of this study is that a complexity analysis of the \ac{SPEM} represents a potential window to evaluate pathological behaviours in the oculomotor system. In addition, a complexity analysis of \ac{SPEM} in \ac{PD} patients is expected to reveal significant information for a proper understanding of the oculomotor mechanisms of Parkinsonian patients. Additionally, according to the theory of complexity loss with ageing and disease progression~\cite{goldberger2002physiologic}, a complexity analysis is expected to derive indicators able to characterise the deviations of the oculomotor system due to the underlying pathophysiological condition introduced by \ac{PD}. 

The literature reports complexity and nonlinear dynamics measures applied to the screening of \ac{PD} using different biometrics, such as \ac{EEG}, voice and gait~\cite{pappalettera2022analysis,moro2021advances,pasluosta2018motor}. Besides, the complexity and nonlinear dynamics of eye movement sequences have been applied to characterise other ocular tasks different from \ac{SPEM}, such as fixations~\cite{harezlak2019understanding}, saccades~\cite{acstefuanoaei2014dynamical}, visual search~\cite{murata2015nonlinear} and accommodation~\cite{hampson2017sensitivity}. However, these measures have received almost no attention to characterise continuous \ac{SPEM} recordings from
participants with \ac{PD}. To our knowledge, only the preliminary study in~\cite{bejani2022detection} has used measures of complexity and nonlinear dynamics extracted from \ac{SPEM} recordings for the automatic screening of \ac{PD}, reporting moderate discrimination capabilities. Thus, evidence remains limited on whether these measures discriminate participants with \ac{PD} from \acp{HC} and about their discriminative information across pursuit tasks and axes.

Regarding the automatic screening of \ac{PD} using  biometrics extracted from eye movements following other feature engineering approaches, the literature is also scarce. In this regard, the work presented in \cite{brien2023classification} recorded an interleaved prosaccade and antisaccade task with a video-based eye tracker to categorise 121 patients with \ac{PD} and 106 \acp{HC}, classifying them from saccade, pupil and blink features with machine learning, obtaining an \ac{AUC} of 0.88. The authors in~\cite{Jiang2025Diagnosis}  recorded gaze stability, prosaccade, antisaccade and \ac{SPEM} tasks in a virtual reality environment from 114 patients with \ac{PD} and 125 \acp{HC}, using a deep learning model to categorise both classes, and reporting an \ac{AUC} of 0.97 in cross-validation. The work in~\cite{Zhang2026VR} also used virtual reality eye tracking with saccade, antisaccade and \ac{SPEM} paradigms from 44 patients with \ac{PD} and 34 \acp{HC}, and a model combining three oculomotor parameters, reporting an \ac{AUC} of 0.93. On the other hand, \cite{Tan2026DPMDLA} analysed smartphone videos of 18 patients with \ac{PD} and classified 2-s windows of horizontal pursuit as normal or abnormal with a deep learning model, thus identifying abnormal pursuit events rather than patients with \ac{PD}. In these studies, \ac{SPEM} were either not used or were one of several tasks, so their specific contribution was not analysed, and the different tasks, technologies, units of analysis and validation designs prevent a direct comparison of their performance.

On the other hand, the strategy followed to preprocess the \ac{SPEM} sequences is also a relevant issue to be studied before any feature extraction as a prior step to extract the most relevant information and to remove potential biases. Blinks produce missing blocks and artificial transitions~\cite{bejani2025imputation}, requiring methods to compensate them. Slow components associated with the presence of a \ac{BW} can modify the temporal and amplitude structure~\cite{Bejani2023BaselineWR}. In addition, differences in the coordination of both eyes can also affect the analysis, which is addressed by estimating the \ac{CERM} component, representing the component common to both eyes~\cite{luque2024estimation}. %
The potential introduction of these preprocessing strategies significantly modifies the characteristics of the sequences obtained, also affecting both their interpretation and the information available to feed a classifier. Thus, a specific study of the effect introduced  by each of these preprocessing strategies is mandatory to fill the associated gap.

This work presents and evaluates an automatic classification pipeline to support \ac{PD} screening from \ac{SPEM} recordings. Complexity and nonlinear dynamics measures are extracted from gaze position sequences preprocessed following different strategies. A common feature subset is selected to feed different \acp{SVM}, each associated with a pursuit task and axis. In addition, their outputs are combined through a \ac{MoE} approach to produce one final score per participant. %
The central question to answer is whether the measures considered provide consistent discrimination between participants with \ac{PD} and \acp{HC}, while the proposed pipeline is intended to support the screening and biomarker research, not to replace clinical diagnosis.

\section*{Materials}

\subsection*{Participants and clinical assessments}

Fifty-two participants with early \ac{PD} and 48 \acp{HC}, matched for age and sex, were recruited. Mean age was 63.80~years in the \ac{PD} group (range: 44--84) and 64.26~years in the \ac{HC} group (range: 46--80).

Participants were required to maintain a stable head position, follow the target throughout the \acp{SPT} and provide recordings that met the technical quality requirements. Twelve participants, comprising 11 with \ac{PD} and one \ac{HC}, were excluded because they could not complete the \acp{SPT} or because their recordings presented quality problems, including double corneal reflections. This left 41 participants with early \ac{PD} and 47 \acp{HC} for the main analysis. The same requirements were applied to both diagnostic groups; they define the population to which the results apply as participants who can complete the \acp{SPT} and provide technically valid recordings.

Data were collected over 2~years at Hospital Universitario de Fuenlabrada and Hospital General Universitario Gregorio Marañón (Community of Madrid, Spain) using the same protocol, equipment and clinical team.

Clinical evaluation included medical history, age at \ac{PD} diagnosis, time since diagnosis, symptoms, complications, the \ac{UPDRS}, the \ac{MoCA} and the Beck Depression Inventory. Table~\ref{tab:participant-clinical-characteristics} summarises the demographic and clinical characteristics. All participants with \ac{PD} were assessed in the ON-medication state after taking their usual antiparkinsonian medication at the normally scheduled time.

\begin{manuscripttable}
\centering
\caption{Demographic and clinical characteristics of the participants included in the main analysis. Continuous variables are mean \(\pm\) standard deviation and male sex is number (percentage). Dashes denote variables specific to participants with \ac{PD}.}
\label{tab:participant-clinical-characteristics}
\uniformtablesetup
\begin{tabularx}{\textwidth}{@{}
>{\raggedright\arraybackslash}X
>{\centering\arraybackslash}m{0.20\textwidth}
>{\centering\arraybackslash}m{0.20\textwidth}
>{\centering\arraybackslash}m{0.20\textwidth}
@{}}
\toprule
\textbf{Variable} &
\makecell[c]{\textbf{All participants}\\\textbf{(\(n=88\))}} &
\makecell[c]{\textbf{\ac{PD}}\\\textbf{(\(n=41\))}} &
\makecell[c]{\textbf{\acp{HC}}\\\textbf{(\(n=47\))}} \\
\midrule
Age, years & \(63.17 \pm 8.56\) & \(62.29 \pm 8.91\) & \(63.96 \pm 8.26\) \\
Male sex, \(n\) (\%) & \(44\) (\(50.00\%\)) & \(24\) (\(58.54\%\)) & \(20\) (\(42.55\%\)) \\
\ac{MoCA} score & \(26.39 \pm 3.04\) & \(26.03 \pm 3.56\) & \(26.70 \pm 2.52\) \\
\ac{UPDRS} score & \(8.01 \pm 10.68\) & \(17.58 \pm 9.38\) & \(0.28 \pm 0.83\) \\
Age at diagnosis, years & --- & \(60.18 \pm 8.36\) & --- \\
Time since diagnosis, years & --- & \(1.73 \pm 0.78\) & --- \\
\bottomrule
\end{tabularx}
\end{manuscripttable}

The ethics committees of both hospitals approved the protocol and all participants provided written informed consent.

\subsection*{Recording and smooth pursuit tasks}

Binocular eye movements were recorded at 1000~Hz with an EyeLink\textsuperscript{®} 1000 Plus infrared video eye tracker (SR Research Ltd, Ontario, Canada). One computer controlled the system and another presented stimuli on an LED monitor 60~cm from the participant. A chin rest was positioned 50~cm from the upper camera knob and the system was calibrated before each session using a nine-point grid.

Participants fixated on a small white central dot for 200~ms before a red target moved along horizontal, vertical or two-dimensional lemniscate trajectories (Fig.~\ref{fig:red-dot}); each trajectory type included four frequency and amplitude combinations (Table~\ref{tab:table1})~\cite{bejani2022detection}. All participants performed 12 \acp{SPT} in a fixed order from \ac{SPT}1 to \ac{SPT}12. During each \ac{SPT}, timestamped X and Y gaze positions were recorded for both eyes. The ordered samples for each eye and axis formed one univariate gaze position sequence, yielding four per recording. The target trajectory denotes the complete spatial path, whereas a target position sequence contains its samples over time along one axis.

\begin{manuscriptfigure}
\centering
\captionsetup[subfigure]{font=footnotesize,skip=2pt}

\begin{subfigure}[t]{0.31\textwidth}
    \centering
    \includegraphics[width=0.85\linewidth]{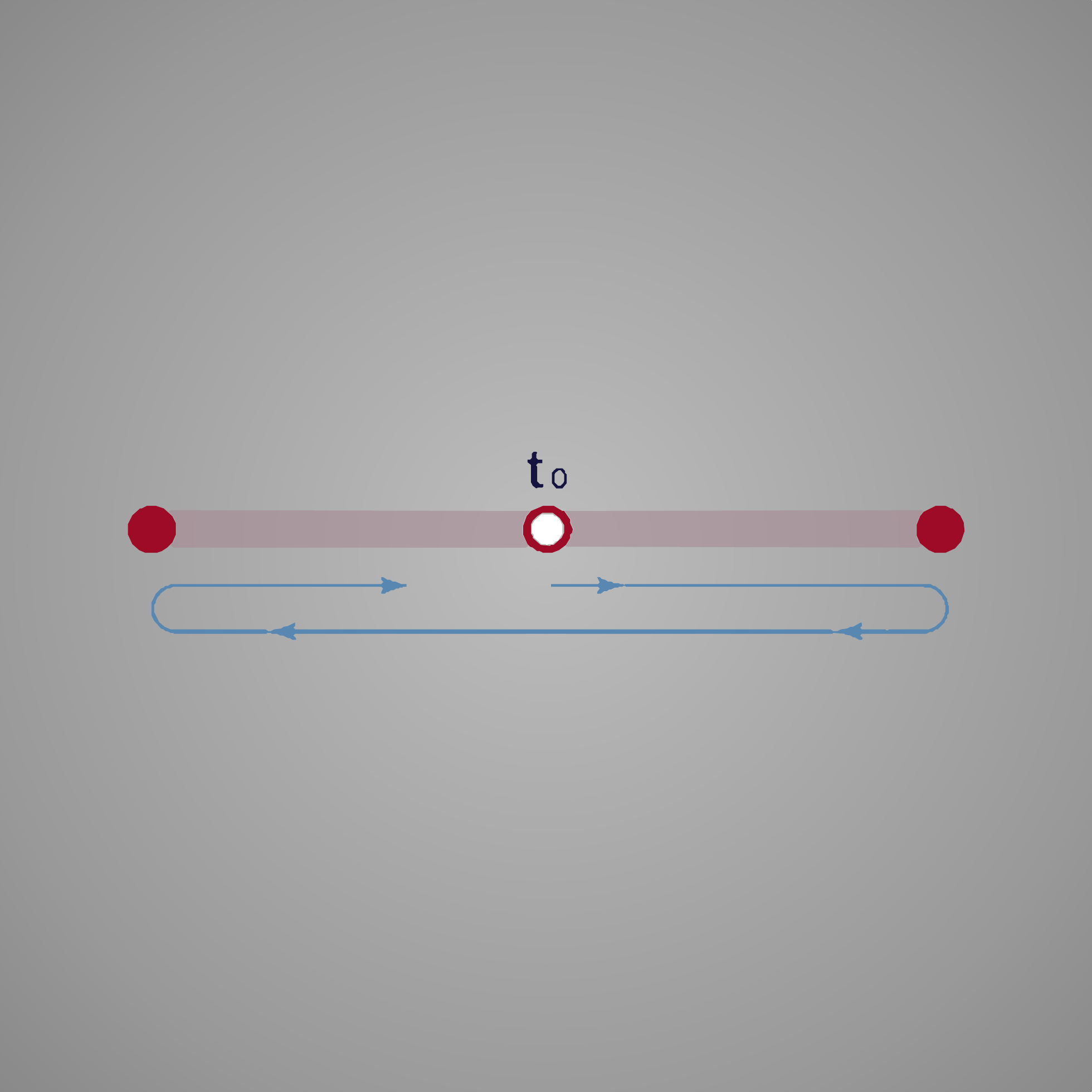}
    \caption{}
    \label{fig:red-dot-a}
\end{subfigure}
\hspace{0.035\textwidth}
\begin{subfigure}[t]{0.31\textwidth}
    \centering
    \includegraphics[width=0.85\linewidth]{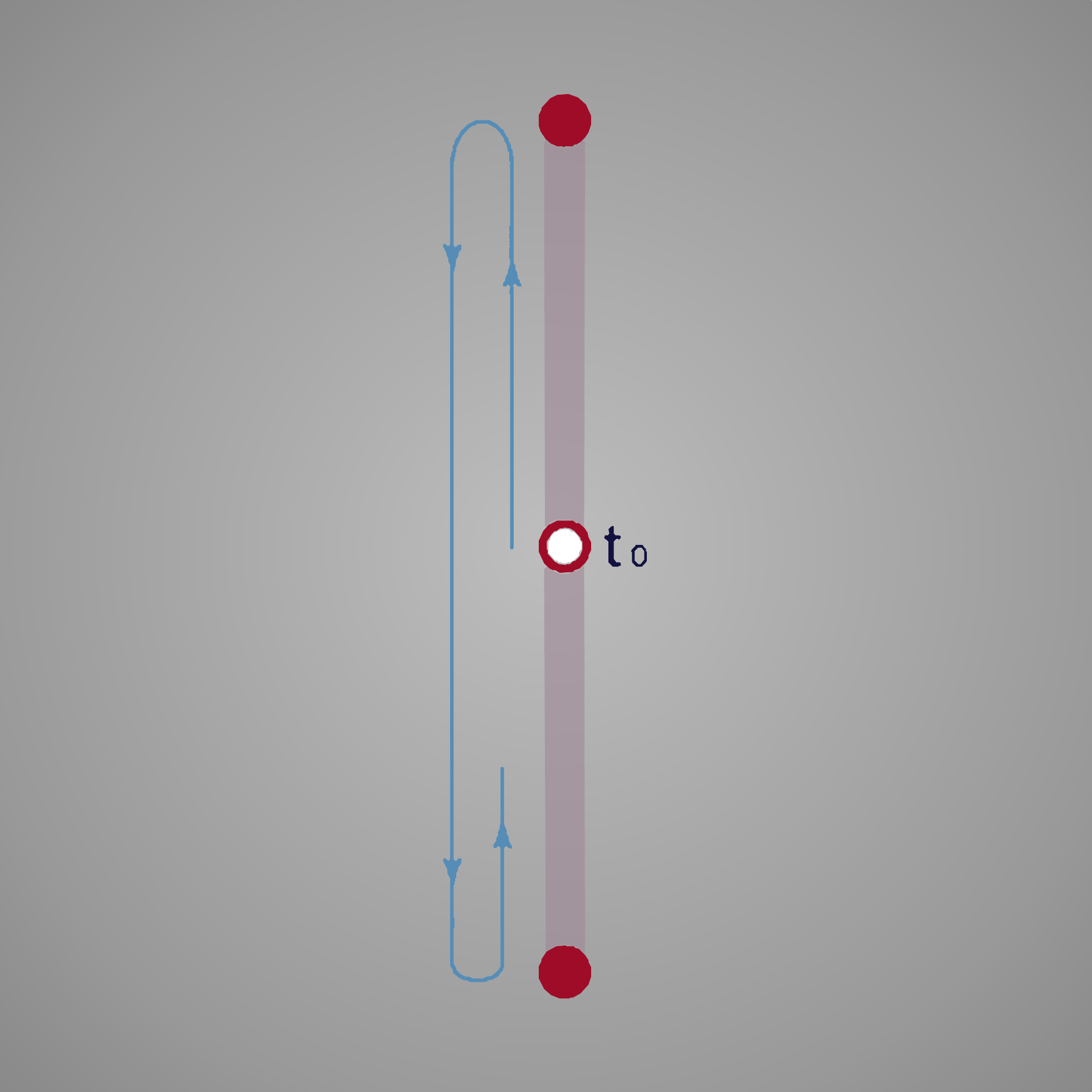}
    \caption{}
    \label{fig:red-dot-b}
\end{subfigure}

\vspace{0.3em}

\begin{subfigure}[t]{0.31\textwidth}
    \centering
    \includegraphics[width=0.85\linewidth]{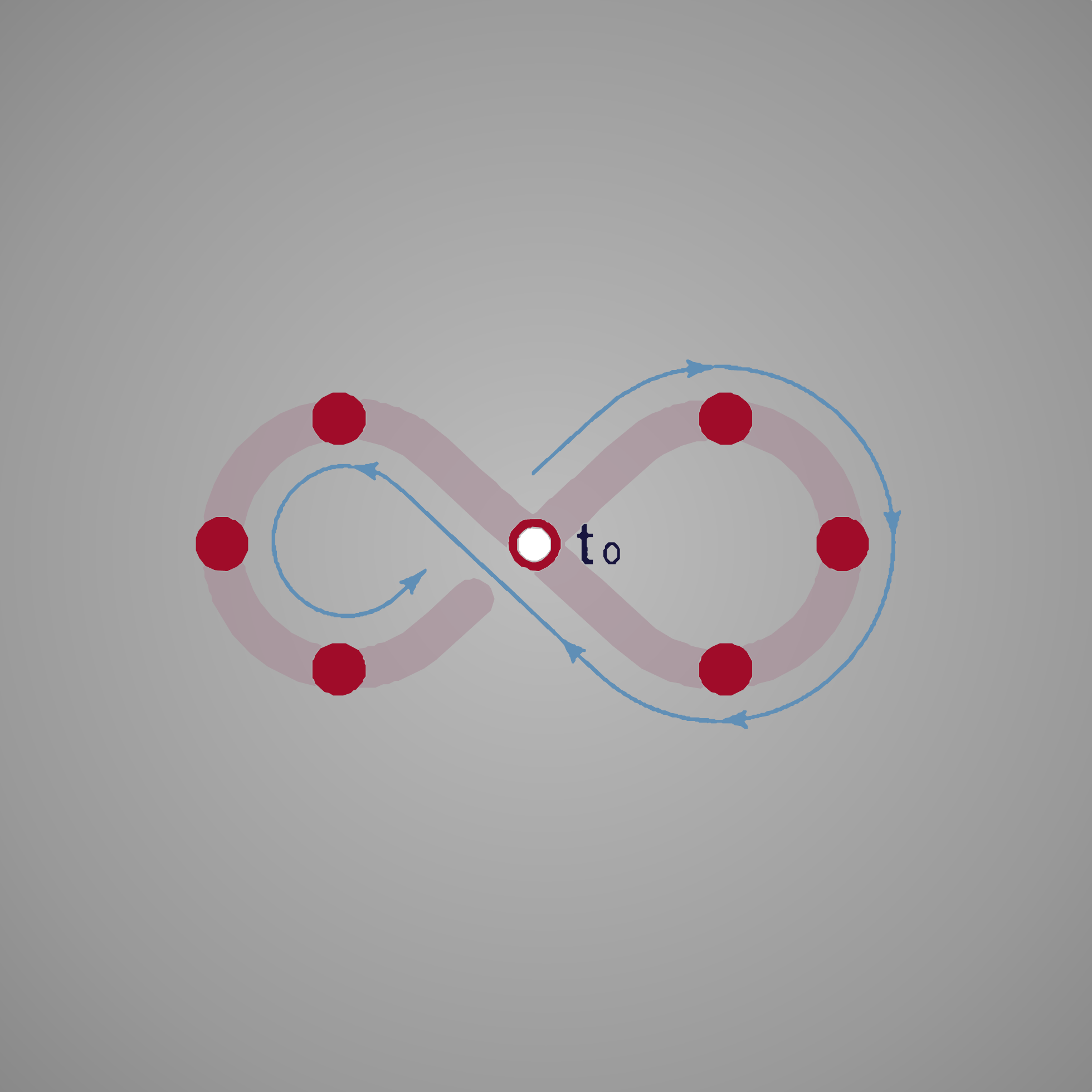}
    \caption{}
    \label{fig:red-dot-c}
\end{subfigure}
\hspace{0.035\textwidth}
\begin{subfigure}[t]{0.31\textwidth}
    \centering
    \includegraphics[width=0.85\linewidth]{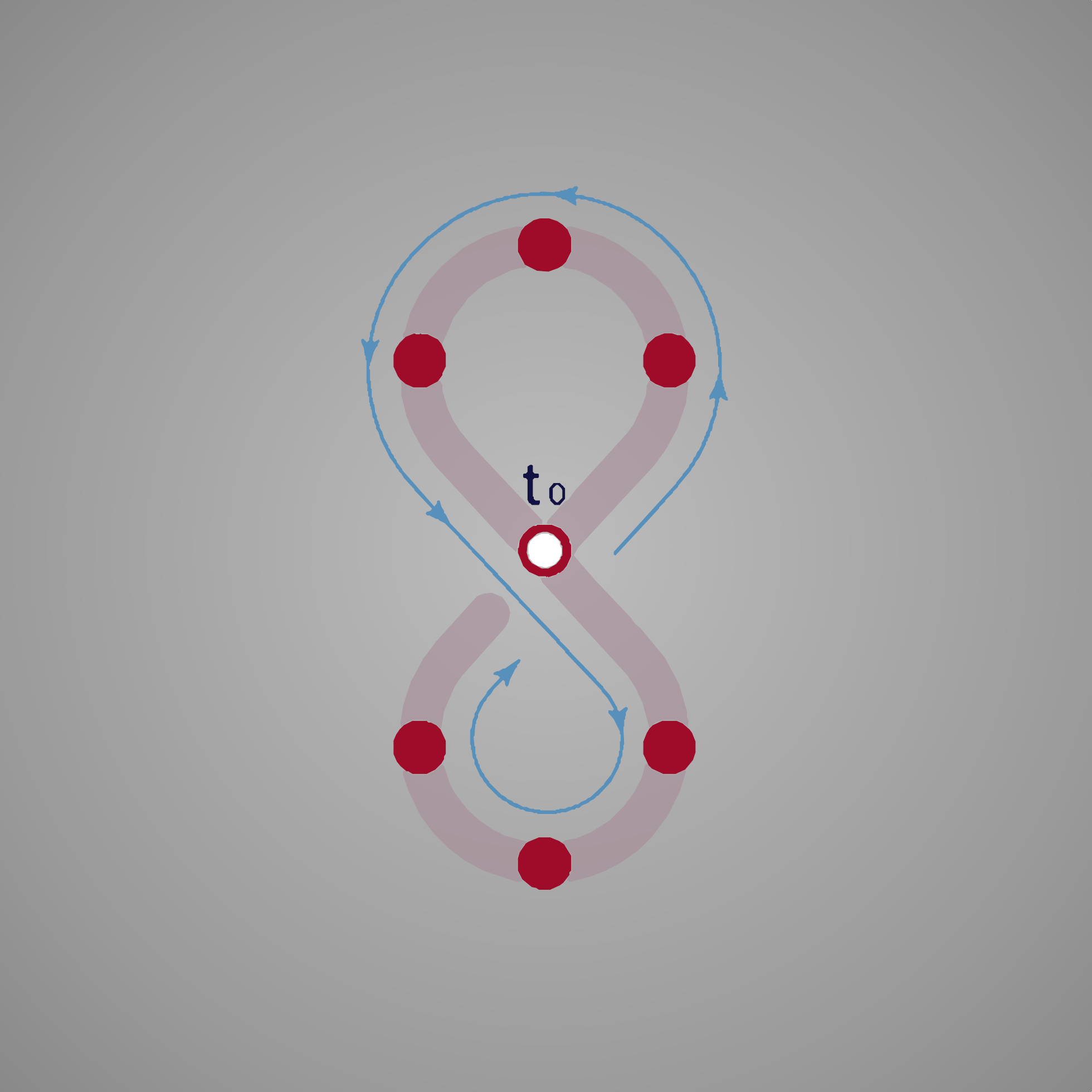}
    \caption{}
    \label{fig:red-dot-d}
\end{subfigure}

\caption{Stimulus trajectories used in the 12 \acp{SPT}: (a) horizontal trajectories used in \ac{SPT}1--\ac{SPT}4; (b) vertical trajectories used in \ac{SPT}5--\ac{SPT}8; (c) horizontal lemniscate trajectories used in \ac{SPT}9 and \ac{SPT}10; and (d) vertical lemniscate trajectories used in \ac{SPT}11 and \ac{SPT}12. Participants were instructed to fixate on the white central point for 200~ms before a red target moved in the direction indicated by the arrows. Target frequency and amplitude varied across \acp{SPT} as detailed in Table~\ref{tab:table1}.}
\label{fig:red-dot}
\end{manuscriptfigure}

\begin{manuscripttable}
\centering
\caption{Target frequency and amplitude for the 12 \acp{SPT}.}
\label{tab:table1}
\uniformtablesetup

\begin{tabular*}{\textwidth}{@{\extracolsep{\fill}}llcccc@{}}
\toprule
\multicolumn{2}{c}{} &
\textbf{\shortstack{X frequency\\(Hz)}} &
\textbf{\shortstack{X amplitude\\($^{\circ}$)}} &
\textbf{\shortstack{Y frequency\\(Hz)}} &
\textbf{\shortstack{Y amplitude\\($^{\circ}$)}} \\
\midrule

\multirow{4}{*}{\textbf{Horizontal}}
& \textbf{\ac{SPT}1} & 0.2 & 6 & 0 & 0 \\
& \textbf{\ac{SPT}2} & 0.2 & 9 & 0 & 0 \\
& \textbf{\ac{SPT}3} & 0.4 & 6 & 0 & 0 \\
& \textbf{\ac{SPT}4} & 0.4 & 9 & 0 & 0 \\
\addlinespace[2pt]

\multirow{4}{*}{\textbf{Vertical}}
& \textbf{\ac{SPT}5} & 0 & 0 & 0.2 & 4 \\
& \textbf{\ac{SPT}6} & 0 & 0 & 0.2 & 7 \\
& \textbf{\ac{SPT}7} & 0 & 0 & 0.4 & 4 \\
& \textbf{\ac{SPT}8} & 0 & 0 & 0.4 & 7 \\
\addlinespace[2pt]

\multirow{4}{*}{\textbf{2D}}
& \textbf{\ac{SPT}9}  & 0.2 & 6 & 0.4 & 3 \\
& \textbf{\ac{SPT}10} & 0.2 & 9 & 0.4 & 6 \\
& \textbf{\ac{SPT}11} & 0.4 & 3 & 0.2 & 6 \\
& \textbf{\ac{SPT}12} & 0.4 & 3 & 0.2 & 7 \\
\bottomrule
\end{tabular*}

\end{manuscripttable}

\section*{Methods}

The development set (70 participants: 33 with \ac{PD} and 37 \acp{HC}) was used to design and train the classification pipeline. The external test set (8 participants with \ac{PD} and 10 \acp{HC}) was used only for its final evaluation and the \ac{YC} cohort (12 participants; mean age 25~years, range 23--30) only to describe its predictions for younger healthy participants.

\subsection*{Preprocessing}

Preprocessing was implemented in MATLAB\textsuperscript{®}. Blink correction was applied to all gaze position sequences, whereas \ac{BW} removal and \ac{CERM} extraction were applied or omitted to define four \acp{EF} (Table~\ref{tab:exp_settings}).

\paragraph{Blink detection and imputation.}
Complete eyelid closure produces missing intervals, often bordered by nonphysiological amplitude changes. An algorithm based on~\cite{luque2024estimation} expanded each interval across these changes until smoother transitions were reached and reconstructed the removed segment with \ac{SSA}~\cite{kondrashov2010gap} (Fig.~\ref{fig:imputation}). The blink-corrected gaze position sequences of the two eyes are shown in Fig.~\ref{fig:eye-sequences}a--b.

\begin{manuscriptfigure}
    \centering
    \includegraphics[width=0.90\linewidth]{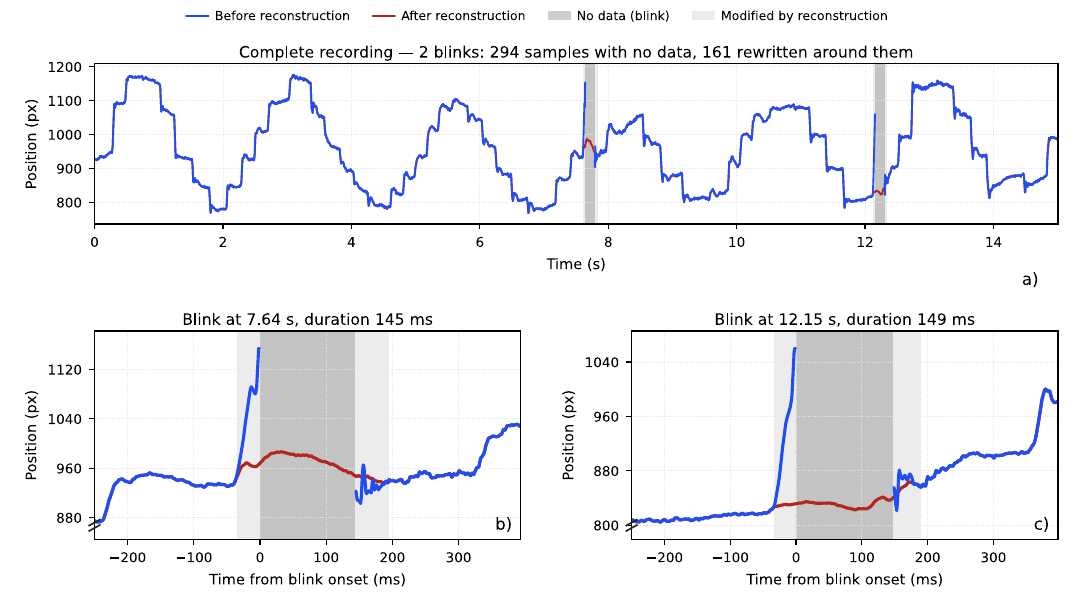}
    \caption{Blink imputation for participant HF011 (\ac{HC}) during \ac{SPT}12, using the gaze position sequence from the left eye along the X axis. Grey regions denote segments reconstructed with \ac{SSA}.}
    \label{fig:imputation}
\end{manuscriptfigure}

\paragraph{Baseline wander removal.}
Calibration, environmental noise, visual impairments, head movement and attention introduce a low-frequency \ac{BW} into the gaze position sequences. \ac{BW} was estimated and removed with a method based on the \ac{EWT}~\cite{Bejani2023BaselineWR} (Fig.~\ref{fig:eye-sequences}c).

\paragraph{Extraction of the \ac{CERM} component.}
\ac{ICA} was applied to the gaze position sequences of the two eyes, before or after \ac{BW} removal, to extract the \ac{CERM} sequence, which is assumed to represent the motor commands common to both eyes~\cite{luque2024estimation} (Fig.~\ref{fig:eye-sequences}d--e).

\begin{manuscriptfigure}
    \centering
    \includegraphics[width=0.90\linewidth]{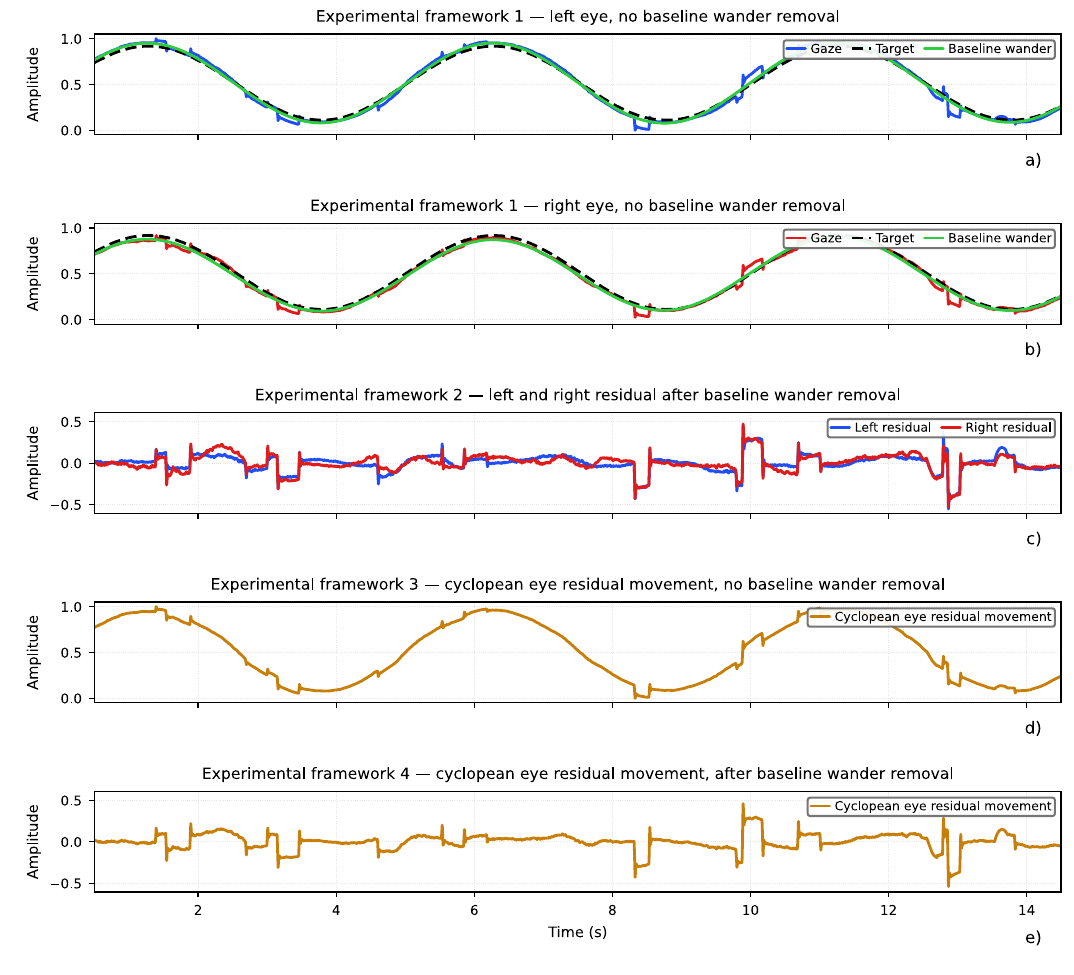}
    \caption{Four \acp{EF} for participant HF009 (\ac{HC}) during \ac{SPT}11 along the X axis: gaze position sequences and estimated \ac{BW} for the left and right eyes before removal (a,b); both gaze position sequences after removal (c); and \ac{CERM} sequences obtained without (d) and with (e) prior \ac{BW} removal.}
    \label{fig:eye-sequences}
\end{manuscriptfigure}

\subsection*{Feature extraction}

Complexity and nonlinear dynamics measures were computed from each gaze position sequence for each participant, \ac{SPT}, eye and axis. A measure is one of the 38 quantities in Table~\ref{tab:table2}, whereas a feature is the value of a measure for one eye and axis in one \ac{SPT}. The measures characterise the temporal and spatial organisation of the sequence rather than its relation to the target position.

\paragraph{Phase space reconstruction.}
\label{sec:phase-space-reconstruction}
Measures requiring an embedding were computed from delay vectors defined by the \ac{EDel} and \ac{EDim}~\cite{packard1980geometry,takens1981detecting}. Four \acp{DS} were compared to select the \ac{EDel}: \ac{DS}1 used a fixed value of 25~samples (25~ms at 1~kHz); \ac{DS}2 used the first local minimum of the \ac{AMI} curve of the target position sequence; and \ac{DS}3 and \ac{DS}4 used the first local minimum and the \(1/e\) decay, respectively, of the \ac{AMI} curve of the gaze position sequence~\cite{fraser1986independent}. \ac{DS}2 gave one \ac{EDel} for each \ac{SPT} and axis, whereas \ac{DS}3 and \ac{DS}4 gave one for each participant. The \ac{EDim} was estimated for each sequence with \ac{FNN}~\cite{kennel1992determining} (Appendix~\ref{app:embedding-parameters}).

\paragraph{Complexity and nonlinear dynamics measures.}
The 38 measures were organised into seven families according to the property they describe (Table~\ref{tab:table2}). \texttt{EntropyHub}\textsuperscript{\textregistered} was used to compute \ac{ApEn}, \ac{SampEn} and the plain, normalised and conditional variants of \ac{FGPE}, \ac{mPE}, \ac{AAPE}, \ac{ePE} and \ac{UQPE}~\cite{Flood2021EntropyHub}. \texttt{NeuroKit2}\textsuperscript{®} was used to compute \ac{ShEn}, \ac{MaxEn}, \ac{DiffEn}, \ac{PowEn}, \ac{TE}, \ac{HEn}, \ac{CEn}, \ac{cApEn}, \ac{QSE}, \ac{PE}, \ac{wPE}, \ac{cPE}, \ac{cwPE}, \ac{cRPE}, \ac{LLE}, \ac{LZC}, \ac{KFD} and \ac{PFD}~\cite{Makowski2021NeuroKit2}. \texttt{Nolds}\textsuperscript{®} was used to compute \ac{HE} and \ac{DFA} and to construct the delay vectors for \ac{CD}~\cite{Scholz2015nolds}.

\newcolumntype{Y}{>{\RaggedRight\arraybackslash}X}

\begin{manuscripttable}
\centering
\caption{Complexity and nonlinear dynamics measures, the sequence property described by each family and whether its measures are computed from delay vectors.}
\label{tab:table2}
\uniformtablesetup
\begin{tabularx}{\textwidth}{@{}p{0.17\textwidth}Yp{0.27\textwidth}p{0.12\textwidth}@{}}
\toprule
\textbf{Family} &
\textbf{Measures} &
\textbf{Sequence property} &
\textbf{Delay vectors} \\
\midrule

Informational entropy &
\ac{ShEn}~\cite{Shannon1948}, \ac{MaxEn}~\cite{Jaynes1957MaxEnt}, \ac{DiffEn}~\cite{Shannon1948}, \ac{PowEn}~\cite{Shannon1948,Makowski2021NeuroKit2}, \ac{TE}~\cite{Tsallis1988}, \ac{HEn}~\cite{Hartley1928}, \ac{CEn}~\cite{Renyi1961} &
Uncertainty and distributional information in sequence amplitudes. &
No \\

Temporal entropy &
\ac{ApEn}~\cite{Pincus1991ApEn}, \ac{cApEn}~\cite{Porta2007cApEn}, \ac{SampEn}~\cite{Richman2000SampEn}, \ac{QSE}~\cite{Lake2006QSE} &
Regularity and recurrence of temporal patterns. &
Yes \\

Ordinal and permutation entropy &
\ac{PE}~\cite{BandtPompe2002PE}, \ac{wPE}~\cite{Fadlallah2013wPE}, \ac{cPE}~\cite{Unakafov2014cPE}, \ac{cwPE}~\cite{Fadlallah2013wPE,Unakafov2014cPE}, \ac{cRPE}~\cite{Renyi1961,Unakafov2014cPE}; plain, normalised and conditional forms of \ac{FGPE}~\cite{LiuWang2009FGPE}, \ac{mPE}~\cite{Bian2012mPE}, \ac{AAPE}~\cite{Azami2016AAPE}, \ac{ePE}~\cite{Huo2019EdgePE} and \ac{UQPE}~\cite{Chen2019ImprovedPE} &
Diversity, transitions, amplitudes and modifications of ordinal patterns. &
Yes \\

Symbolic complexity &
\ac{LZC}~\cite{lempel1976complexity} &
Novelty rate of symbolic patterns. &
Yes \\

Geometric and fractal &
\ac{KFD}~\cite{katz1988fractals}, \ac{PFD}~\cite{petrosian1995kolmogorov} &
Geometric waveform complexity. &
No \\

Phase space dynamics &
\ac{LLE}~\cite{wolf1985determining}, \ac{CD}~\cite{GrassbergerProcaccia1983CD} &
Divergence and dimensionality of reconstructed trajectories. &
Yes \\

Memory and scaling &
\ac{HE}~\cite{hurst1951long}, \ac{DFA}~\cite{peng1994mosaic} &
Long-range dependence and scaling behaviour. &
No \\

\bottomrule
\end{tabularx}
\end{manuscripttable}

\subsection*{Experimental setup}

The four \acp{EF} and the four \acp{DS} were combined into 16 experimental settings (Table~\ref{tab:exp_settings}), so that preprocessing and delay strategy could be compared separately.

\begin{manuscripttable}
\centering
\caption{The 16 experimental settings formed by the four \acp{EF} and four \acp{DS}.}
\label{tab:exp_settings}

\begin{tabular*}{\textwidth}{@{\extracolsep{\fill}}lcccc@{}}
\toprule

\shortstack[c]{
\textbf{Delay Strategy} \(\rightarrow\)\\
\textbf{Experimental Framework} \(\downarrow\)
} &
\shortstack{
\textbf{\ac{DS}1}\\
\scriptsize Fixed delay\\
\scriptsize 25~samples
} &
\shortstack{
\textbf{\ac{DS}2}\\
\scriptsize Target position \ac{AMI}\\
\scriptsize first minimum
} &
\shortstack{
\textbf{\ac{DS}3}\\
\scriptsize Gaze position \ac{AMI}\\
\scriptsize first minimum
} &
\shortstack{
\textbf{\ac{DS}4}\\
\scriptsize Gaze position \ac{AMI}\\
\scriptsize \(1/e\) decay
} \\

\midrule

\shortstack[l]{
\textbf{\ac{EF}1}\\
\scriptsize No \ac{BW} removal\\
\scriptsize No \ac{CERM} extraction
} &
\ac{EF}1--\ac{DS}1 &
\ac{EF}1--\ac{DS}2 &
\ac{EF}1--\ac{DS}3 &
\ac{EF}1--\ac{DS}4 \\

\shortstack[l]{
\textbf{\ac{EF}2}\\
\scriptsize \ac{BW} removal\\
\scriptsize No \ac{CERM} extraction
} &
\ac{EF}2--\ac{DS}1 &
\ac{EF}2--\ac{DS}2 &
\ac{EF}2--\ac{DS}3 &
\ac{EF}2--\ac{DS}4 \\

\shortstack[l]{
\textbf{\ac{EF}3}\\
\scriptsize No \ac{BW} removal\\
\scriptsize \ac{CERM} extraction
} &
\ac{EF}3--\ac{DS}1 &
\ac{EF}3--\ac{DS}2 &
\ac{EF}3--\ac{DS}3 &
\ac{EF}3--\ac{DS}4 \\

\shortstack[l]{
\textbf{\ac{EF}4}\\
\scriptsize \ac{BW} removal\\
\scriptsize \ac{CERM} extraction
} &
\ac{EF}4--\ac{DS}1 &
\ac{EF}4--\ac{DS}2 &
\ac{EF}4--\ac{DS}3 &
\ac{EF}4--\ac{DS}4 \\

\bottomrule
\end{tabular*}
\end{manuscripttable}

\subsection*{Preliminary selection of the experimental setting and organisation of the base models}
\label{sec:experimental_setting_selection}

The development set was used to select the \ac{DS}, compare the \acp{EF} and define the organisation of the base models, that is, the \acp{SVM} whose outputs are combined in the \ac{MoE}. All comparisons used the same logistic regression classifier (\(\ell_2\) regularisation, \(C=1\)), so that classifier optimisation did not confound them. Performance was estimated by \ac{LOPO} \ac{CV}, with standardisation fitted within each training fold and \ac{AUC} calculated from the pooled out-of-fold predictions~\cite{varma2006bias}.

The \acp{DS} were compared by \ac{AUC} using all features, the features computed from delay vectors and the features computed directly from the sequence; only the second set depends on the \ac{EDel} and \ac{EDim}. To interpret the temporal scale captured by the features computed from delay vectors, the temporal span of the delay vectors was calculated for each \ac{DS}:

\begin{equation}
T_{\mathrm{span}}=(m-1)\tau,
\label{eq:temporal-span}
\end{equation}

where \(m\) is the \ac{EDim} and \(\tau\) the \ac{EDel}~\cite{takens1981detecting}. With the selected \ac{DS}, three organisations of the base models were compared by \ac{AUC}: one base model for each \ac{SPT}, pooling both axes and both eyes (12 base models); one base model for each \ac{SPT} and axis, pooling both eyes (24 base models); and one base model for each \ac{SPT}, axis and eye (48 base models). The four \acp{EF} were then compared by \ac{AUC} with all features and by Cohen's \(d\), to check whether the direction of each family was preserved~\cite{lakens2013calculating}.

To interpret differences between \acp{EF}, the effect of \ac{BW} removal and \ac{CERM} extraction on the sequences was described in terms of global dispersion and local variation between consecutive samples. For \ac{BW} removal, global dispersion was measured by the standard deviation of the sequence and local variation by the standard deviation of its first differences, \(\Delta x_t=x_t-x_{t-1}\). Because the scale and sign of the components extracted by \ac{ICA} are arbitrary, each \ac{CERM} sequence is min-max normalised to \([0,1]\)~\cite{luque2024estimation}, whereas the gaze position sequences retain the units of the eye tracker, recorded with the same geometry for all participants. The amplitude of the two cannot be compared directly, so \ac{CERM} extraction was assessed by the relative local variation, defined as the standard deviation of the first differences divided by the standard deviation of the sequence. It was calculated for the complete sequence and in 50~ms windows with \(50\%\) overlap. The values from the two eyes were averaged before comparison with the \ac{CERM} sequence. For each participant, \ac{SPT} and axis, the change in each statistic \(M\) was expressed as \(R=M_{\mathrm{after}}/M_{\mathrm{before}}\), so that \(R<1\) indicates a reduction. Ratios were summarised within participants and diagnostic group medians with \(95\%\) confidence intervals were estimated from \(10\,000\) bootstrap repetitions that resampled participants. These analyses were descriptive and were not used to select the \ac{EF}.

The selected experimental setting and organisation of the base models were fixed before feature selection and model selection.

\subsection*{Classification pipeline and validation strategy}

Each base model was an \ac{SVM} trained on the features of one \ac{SPT}, and all base models were combined into an \ac{MoE} (Fig.~\ref{fig:study-pipeline}), in which they play the role usually referred to as experts~\cite{jacobs1991adaptive}. The assignment of axes and eyes to base models was fixed by the preliminary analysis described above. All base models shared the same feature subset and \ac{SVM} configuration. Selecting them separately for each base model would multiply the choices fitted to 70 participants and increase the risk of overfitting. The pipeline therefore required three choices: the feature subset, the \ac{SVM} configuration and the combiner. They were made using 20 repeated holdouts of the development set, each with 56 training and 14 validation participants and with all recordings of a participant in the same partition. The pipeline was implemented with \texttt{scikit-learn}\textsuperscript{®}~\cite{pedregosa2011scikit}.

\begin{pipelinefigure}
\centering
\includegraphics[width=\linewidth]{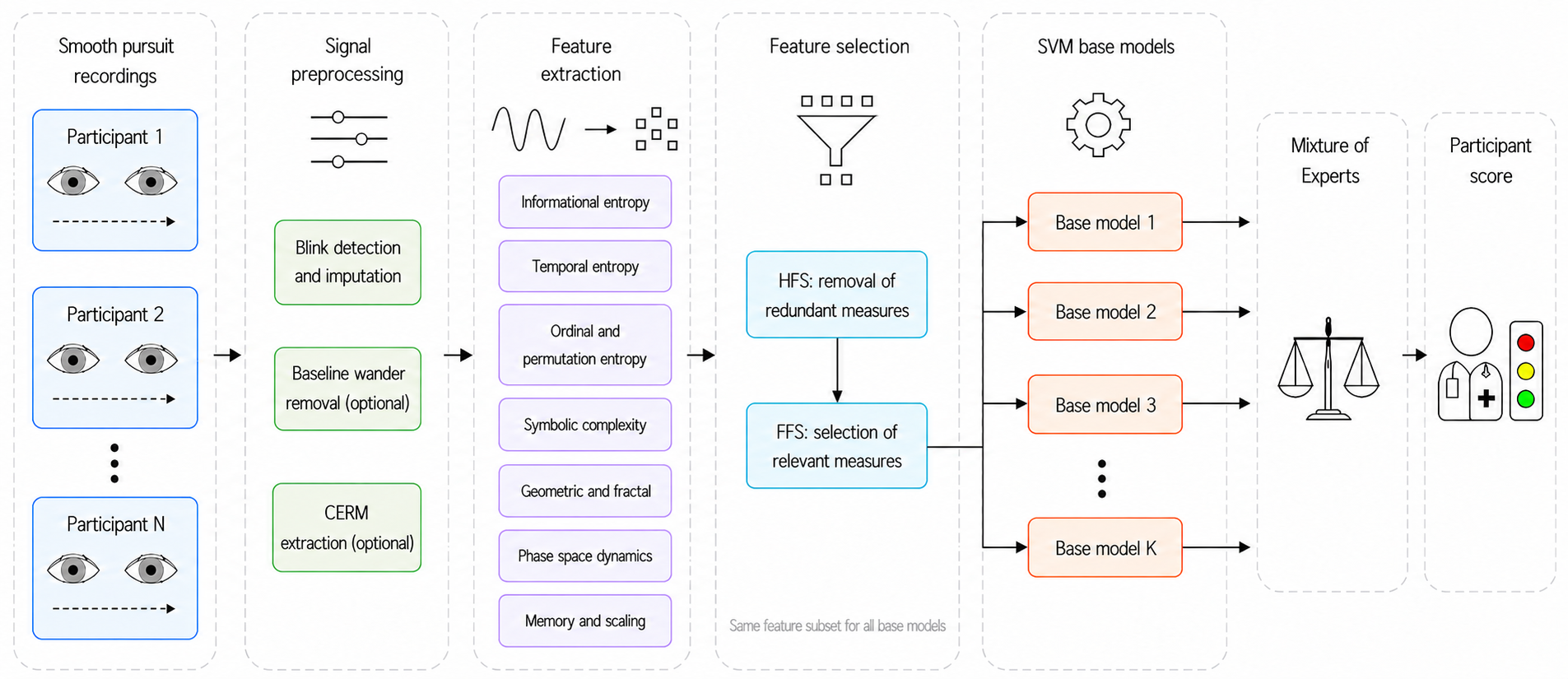}
\pipelinecaption{Overview of the pipeline: from preprocessing and feature extraction to participant-level classification and external evaluation.}
\label{fig:study-pipeline}
\end{pipelinefigure}

Every choice was evaluated with the complete \ac{MoE}, because participants are classified by the combined output rather than by individual base models. The criterion was the geometric mean of sensitivity and specificity (G-mean) on the validation participants, which is given by: 

\begin{equation}
G=\sqrt{\mathrm{sensitivity}\times\mathrm{specificity}},
\label{eq:gmean}
\end{equation}

which is high only when both diagnostic groups are correctly classified~\cite{kubat1997addressing}. Configurations were ranked by:

\begin{equation}
\begin{aligned}
S_{\text{gen}}
&=\overline{G}_{\text{val}}
-0.25\,\sigma_{G,\text{val}} \\
&\quad -0.5\,\max\!\left(
\overline{G}_{\text{train}}
-\overline{G}_{\text{val}},0
\right),
\end{aligned}
\label{eq:generalisation-score}
\end{equation}

which penalises variability across holdouts and positive training--validation gaps, so that stable configurations that do not overfit rank higher~\cite{cawley2010overfitting}.

The \ac{SVM} hyperparameter search covered \(C\), kernel type, \(\gamma\) for the \ac{RBF} kernel and class weight, with logarithmic grids for \(C\) and \(\gamma\)~\cite{hsu2003practical}. Class weights tested whether a greater cost for misclassifying \ac{PD} improved classification~\cite{akbani2004applying}:

\begin{equation}
\begin{aligned}
C &\in \{0.1,\,1,\,10,\,100\}, \\
\text{kernel} &\in \{\text{linear},\,\text{rbf}\}, \\
\gamma &\in \{0.01,\,0.1,\,1.0\}
\quad (\ac{RBF}\text{ kernel only}), \\
\text{class\_weight} &\in
\{\text{none},\,\{0\!:\!1,1\!:\!2\},\,\{0\!:\!1,1\!:\!5\}\}.
\end{aligned}
\label{eq:svm-search-space}
\end{equation}

Each of the 48 \ac{SVM} configurations was evaluated jointly with each of the eight combiner configurations described below.

\subsubsection*{Feature selection}

Feature selection combined two stages: \ac{HFS} removes redundant measures without class labels and \ac{FFS} retains the relevant ones~\cite{mitra2002feature,kohavi1997wrappers}. Both stages operated on measures, so that each retained measure provided its features for every eye and axis of the base models. To assess the contribution of each stage, the pipeline was also evaluated with the complete feature set and with the \ac{RFS}, formed by the features of the representative measures obtained with \ac{HFS}.

\paragraph{Hierarchical Feature Selection.}
Many measures, such as the variants of \ac{PE}, are alternative definitions of the same quantity and are strongly correlated. When features are nearly collinear, the weight assigned to each one is not identifiable and its importance depends on noise~\cite{buhlmann2013correlated}. \ac{HFS} removes this redundancy by grouping strongly correlated measures and keeping one measure per group.

Two measures were considered redundant when their absolute Spearman correlation was at least \(0.80\) in at least \(80\%\) of the 48 combinations of \ac{SPT}, eye and axis, so that redundancy held across recording conditions. Spearman correlation was used to capture monotonic, not only linear, relations. To ensure that redundancy did not depend on particular participants, this criterion was evaluated in repeated subsamples of the development participants~\cite{kalousis2007stability,nogueira2018stability}. For each pair of measures \(i\) and \(j\), the consensus \(c_{ij}\) was the proportion of subsamples in which the pair was redundant; \(c_{ij}=1\) indicates that the two measures were redundant in every subsample. Measures were then grouped so that every pair within a group had \(c_{ij}\geq0.80\). This grouping was obtained by agglomerative hierarchical clustering with distance \(1-c_{ij}\) and complete linkage, which constrains every pair within a group rather than their average~\cite{johnson1967hierarchical}.

Each group was represented by the member most strongly associated with the rest of the group~\cite{kaufman1990findinggroups}. Keeping an original measure rather than a combination such as a principal component preserves its interpretation. Because \ac{HFS} does not use class labels, it was applied once to the complete development set without biasing the holdout estimates~\cite{ambroise2002selectionbias}.

\paragraph{Forward Feature Selection.}
A measure without redundant copies may still carry no discriminative information. Uninformative features increase the variance of the classifier without improving discrimination, which raises the risk of overfitting in a small sample~\cite{guyon2003introduction}. \ac{FFS} retains the relevant representatives as a wrapper method~\cite{kohavi1997wrappers}. Starting from an empty subset, each step added the representative that most increased the G-mean of the complete \ac{MoE}, so that measures were evaluated in combination. \ac{FFS} was applied to the \ac{RFS} to avoid arbitrary choices between nearly identical measures.

Forward selection is unstable~\cite{kalousis2007stability,nogueira2018stability}, so \ac{FFS} was repeated on 10 inner partitions of the training participants of each holdout, giving 200 runs. A representative was retained when it was selected in more than a fraction \(f\) of them~\cite{meinshausen2010stability}. The value of \(f\) was chosen within the holdouts by the G-mean of the resulting subsets. The \ac{SVM} configuration and combiner were then selected again with the final subset. To assess whether \ac{FFS} selected better than chance, the final subset was compared with 20 random subsets with the same number of representatives and fitted with the same procedure.

\subsubsection*{Mixture of Experts combination and final evaluation}

Each base model produced one \ac{SVM} decision score per participant, which was converted into a probability using Platt scaling~\cite{platt2000probabilistic}. Three combiners of increasing flexibility were compared: the simple mean, the mean weighted by \ac{AUC} and logistic regression stacking~\cite{kittler1998combining,wolpert1992stacked}. The simple mean fits no parameters, the weighted mean favours the most discriminative base models and logistic regression stacking learns the weight of each base model from the training participants. The product rule, which multiplies the probabilities of the base models or, equivalently, sums their logarithms, was not considered because it assumes that the base models are conditionally independent given the class, which does not hold for base models computed from recordings of the same participant~\cite{kittler1998combining}.

Let \(p_e\) be the calibrated probability from base model \(e\) and \(E\) the number of base models. Averaging compresses probabilities towards \(0.5\), so the two means rank the participants but are not calibrated probabilities. Each mean, denoted by \(s\), was therefore mapped to a probability with a logistic recalibration~\cite{cox1958two}. The simple mean is given by:

\begin{equation}
\begin{aligned}
s_{\mathrm{m}}
&=
\frac{1}{E}
\sum_{e=1}^{E} p_e, \\
p_{\mathrm{m}}
&=
\sigma\left(a_{\mathrm{m}}\,\operatorname{logit}(s_{\mathrm{m}})+b_{\mathrm{m}}\right).
\end{aligned}
\label{eq:mean-pooling}
\end{equation}

where \(\sigma(\cdot)\) is the logistic sigmoid, \(\operatorname{logit}(\cdot)\) its inverse, and the slope \(a_{\mathrm{m}}\) and intercept \(b_{\mathrm{m}}\) 
were estimated from out-of-fold scores of the training participants of each holdout.

The weighted mean uses the \ac{AUC} of each base model, calculated within the training participants of each holdout:

\begin{equation}
\begin{aligned}
w_e
&=
\frac{\mathrm{AUC}_e}{\sum_{j=1}^{E}\mathrm{AUC}_j}, \\
s_{\mathrm{w}}
&=
\sum_{e=1}^{E}w_e\,p_e, \\
p_{\mathrm{w}}
&=
\sigma\left(a_{\mathrm{w}}\,\operatorname{logit}(s_{\mathrm{w}})+b_{\mathrm{w}}\right).
\end{aligned}
\label{eq:auc-weighted-pooling}
\end{equation}

where the slope \(a_{\mathrm{w}}\) and intercept \(b_{\mathrm{w}}\) 
were also estimated from out-of-fold scores of the training participants of each holdout.

Logistic regression stacking is based on a logistic regression model applied to the predictions of the base models, which acts as a meta-model that combines them into a single probability. It is defined as:

\begin{equation}
p_{\mathrm{LR}}
=
\sigma\left(
\beta_0+\sum_{e=1}^{E}\beta_e p_e
\right),
\label{eq:logistic-combiner}
\end{equation}

where \(p_{\mathrm{LR}}\) is the output probability of the combiner, \(\beta_0\) the intercept and \(\beta_e\) the coefficient of base model \(e\). As \(p_{\mathrm{LR}}\) is already a probability, it was used without recalibration. The search space of this combiner was

\begin{equation}
\begin{aligned}
C &\in \{0.1,\,1.0,\,10\}, \\
\text{penalty} &\in \{\ell_2,\,\ell_1\}, \\
\text{class\_weight} &\in \{\text{none}\}.
\end{aligned}
\label{eq:logistic-combiner-search-space}
\end{equation}

After model selection, the combiner and calibration model parameters (weights of the base models and, where applicable, regression intercepts and calibration slopes) were averaged across the 20 holdouts to obtain a single calibrated combiner. Development performance was reported as the mean (standard deviation) of the performance metrics obtained by applying this combiner to the validation participants in each holdout. The decision threshold was selected on the resulting 280 validation predictions rather than on the training predictions, because thresholds tuned on the training data are optimistic and may not transfer to new participants~\cite{esposito2021ghost}. It was set at the \ac{EER} operating point, at which the false positive and false negative rates are equal~\cite{martin1997det}, so that both errors were weighted equally, as in the G-mean.

For external evaluation, the base models were refitted on the complete development set, and their predictions were combined and calibrated using these averaged parameters, with the same decision threshold. The pipeline was then evaluated once on the external test set. Confidence intervals of \(95\%\) were obtained from \(10\,000\) bootstrap resamples stratified by diagnostic group for \ac{AUC} and G-mean and from Wilson intervals for sensitivity and specificity~\cite{wilson1927probable}.

Contributions were computed in the development participants and expressed as a percentage of the total. The contribution of each base model was its mean absolute contribution to the output of the combiner. The contribution of each feature within a base model was its mean absolute SHAP value, which, unlike the \ac{SVM} coefficients, is also defined for the \ac{RBF} kernel~\cite{lundberg2017unified}, weighted by the contribution of the base model so that it referred to the final prediction. The direction of a feature was the sign of the association between its values and its SHAP values and was compared with the sign of Cohen's \(d\) between participants with \ac{PD} and \acp{HC}~\cite{lakens2013calculating}.

The same pipeline was applied to the \ac{YC} cohort to compare its probabilities with those of the external \acp{HC} and participants with \ac{PD}.

\section*{Results}

\subsection*{Selection of the experimental setting and organisation of the base models}

The preliminary analyses selected \ac{EF}2--\ac{DS}1 and one base model for each \ac{SPT} and axis, which received the features of each measure from both eyes (Table~\ref{tab:experimental-setting-summary}). The complete comparison of \acp{DS} is reported in Supplementary Table~\ref{tab:supp-delay-strategy-selection}. Pooling both eyes and both axes gave an \ac{AUC} of \(0.66\). Separating the axes while keeping both eyes pooled yielded performance close to chance for the Y axis and a slightly higher \ac{AUC} of \(0.68\) for the X axis compared with pooling both axes. Further separating the eyes within each axis yielded similar values for the left and right eyes along the X axis, both close to that obtained with both eyes pooled, while Y axis performance remained close to chance for either eye (Table~\ref{tab:experimental-setting-summary}; Supplementary Table~\ref{tab:supp-channel-decomposition}).

\begin{manuscripttable}
\centering
\caption{Development set evidence used to select the experimental setting and organisation of the base models.}
\label{tab:experimental-setting-summary}
\uniformtablesetup

\begin{tabularx}{\textwidth}{@{}
>{\RaggedRight\arraybackslash}p{0.16\textwidth}
>{\RaggedRight\arraybackslash}p{0.30\textwidth}
>{\RaggedRight\arraybackslash}X
@{}}
\toprule
\textbf{Choice} &
\textbf{Selected option} &
\textbf{Main quantitative evidence} \\
\midrule

Delay strategy &
\textbf{\ac{DS}1} &
\ac{AUC}: \(0.68 \pm 0.07\) with all features and \(0.65 \pm 0.14\) with features computed from delay vectors; temporal span of the delay vectors: \(50\)--\(75\)~ms
\\

\addlinespace[7pt]

Organisation of the base models &
One base model for each \ac{SPT} and axis, using features from both eyes; \(24\) base models in total &
\ac{AUC}: \(0.68 \pm 0.07\) for the X axis and \(0.51 \pm 0.06\) for the Y axis; \(0.68 \pm 0.07\) for the left eye and \(0.70 \pm 0.08\) for the right eye along the X axis
\\

\addlinespace[7pt]

Experimental framework &
\textbf{\ac{EF}2}
&
\ac{AUC}: \(0.76\) for \ac{EF}2, \(0.71\) for \ac{EF}1, \(0.66\) for \ac{EF}4 and \(0.60\) for \ac{EF}3
\\

\bottomrule
\end{tabularx}
\end{manuscripttable}

\ac{BW} removal reduced the standard deviation of the sequence by \(83.00\%\) and the standard deviation of its first differences by \(3.00\%\). \ac{CERM} extraction reduced the relative local variation, with median ratios in 50~ms windows of \(0.81\) for \ac{EF}3/\ac{EF}1 and \(0.89\) for \ac{EF}4/\ac{EF}2 (Supplementary Table~\ref{tab:supp-framework-sequence-properties}). In the effect size analysis, \ac{BW} removal reduced the absolute Cohen’s \(d\) of \ac{PowEn}, which depends on amplitude, whereas that of \ac{SampEn} changed little. Following \ac{CERM} extraction, differences were observed in both effect magnitude and direction, including reversals for \ac{PowEn} and \ac{SampEn}. Other measures, such as \ac{DFA}, \ac{HE}, \ac{PFD} and \ac{LLE}, showed changes in magnitude while retaining their effect direction across the four \acp{EF} (Supplementary Table~\ref{tab:supp-family-level-framework-selection}).

\subsection*{Feature selection}

\paragraph{Hierarchical Feature Selection.}
\ac{HFS} reduced the 38 measures to 20 representatives. The largest groups contained the nine variants of \ac{PE} and five informational entropies, and \ac{DiffEn} was grouped with \ac{PowEn}. The median absolute Spearman correlation was \(0.96\) between measures of the same group and \(0.31\) between measures of different groups.

Removing this redundancy changed the distribution of the contribution. With the complete feature set, contribution was spread almost evenly among the members of each group, and \(10.61\%\) of the pairs of features from measures of the same group were used in opposite directions. The features of measures in redundancy groups collectively accounted for \(51.22\%\) of the total contribution before redundancy removal and \(22.21\%\) afterwards, when only their representatives were retained. Conversely, the features of measures without redundant partners increased their combined contribution from \(48.78\%\) to \(77.79\%\).

\paragraph{Forward Feature Selection.}
\ac{FFS} retained 13 out of 20 representatives, those selected in more than \(40.00\%\) of the 200 runs: \ac{HE}, \ac{wPE}, \ac{LZC}, \ac{FGPE}, normalised \ac{FGPE}, \ac{cwPE}, conditional \ac{UQPE}, conditional \ac{AAPE}, \ac{DiffEn}, \ac{CD}, \ac{KFD}, \ac{PFD} and \ac{cApEn}. Their development G-mean (\(0.83\)) was higher than that of all 20 random subsets of 13 representatives (mean \(0.78\)).

Table~\ref{tab:performance-summary} compares the three feature sets. The \ac{RFS} maintained the development performance of the complete feature set with 20 measures. \ac{HFS}+\ac{FFS} obtained the highest mean \ac{AUC} and G-mean and the lowest G-mean variability with 13 measures.

\begin{manuscripttable}
\centering
\caption{Participant-level performance of the three feature sets. Numbers in parentheses are measures; each measure provided one feature per eye. Development values are the mean (standard deviation) across the 20 holdouts. The external test set was evaluated with the selected feature set, shown in bold.}
\label{tab:performance-summary}
\uniformtablesetup
\begin{tabularx}{\textwidth}{@{}>{\RaggedRight\arraybackslash}X
>{\centering\arraybackslash}m{0.17\textwidth}
>{\centering\arraybackslash}m{0.17\textwidth}
>{\centering\arraybackslash}m{0.17\textwidth}
>{\centering\arraybackslash}m{0.17\textwidth}@{}}
\toprule
& \multicolumn{3}{c}{\textbf{Development set}} & \textbf{External test set} \\
\cmidrule(lr){2-4}\cmidrule(l){5-5}
\textbf{Metric} &
\makecell[c]{\textbf{All features}\\\textbf{(38)}} &
\makecell[c]{\textbf{\acs{RFS}}\\\textbf{(20)}} &
\makecell[c]{\textbf{\acs{HFS}+\acs{FFS}}\\\textbf{(13)}} &
\makecell[c]{\textbf{\acs{HFS}+\acs{FFS}}\\\textbf{(13)}} \\
\midrule
\ac{AUC}    & \(0.88\;(0.05)\) & \(0.90\;(0.08)\) & \(\mathbf{0.91\;(0.06)}\) & \(\mathbf{0.86}\) \\
Sensitivity & \(0.78\;(0.17)\) & \(0.78\;(0.18)\) & \(\mathbf{0.84\;(0.11)}\) & \(\mathbf{0.75}\) \\
Specificity & \(0.78\;(0.18)\) & \(0.78\;(0.21)\) & \(\mathbf{0.84\;(0.16)}\) & \(\mathbf{0.80}\) \\
G-mean      & \(0.76\;(0.10)\) & \(0.77\;(0.15)\) & \(\mathbf{0.83\;(0.06)}\) & \(\mathbf{0.77}\) \\
\bottomrule
\end{tabularx}
\end{manuscripttable}

\subsection*{Performance of the selected pipeline}

The joint search selected an \ac{SVM} with an \ac{RBF} kernel (\(C=1\), \(\gamma=0.01\) and a weight of 5 for the \ac{PD} class) and logistic regression stacking with \(\ell_1\) regularisation and \(C=1\).

Table~\ref{tab:performance-summary} reports the development performance with a decision threshold of \(0.49\). On the external test set, the \ac{AUC} was \(0.86\) (95\% confidence interval: \(0.65\)--\(1.00\)). Six out of 8 participants with \ac{PD} and 8 out of 10 \acp{HC} were correctly classified, giving a sensitivity of \(0.75\) (\(0.41\)--\(0.93\)), a specificity of \(0.80\) (\(0.49\)--\(0.94\)) and a G-mean of \(0.77\) (\(0.55\)--\(0.95\)) (Fig.~\ref{fig:selected-configuration-performance}).

\begin{manuscriptfigure}
\centering
\includegraphics[width=\linewidth]{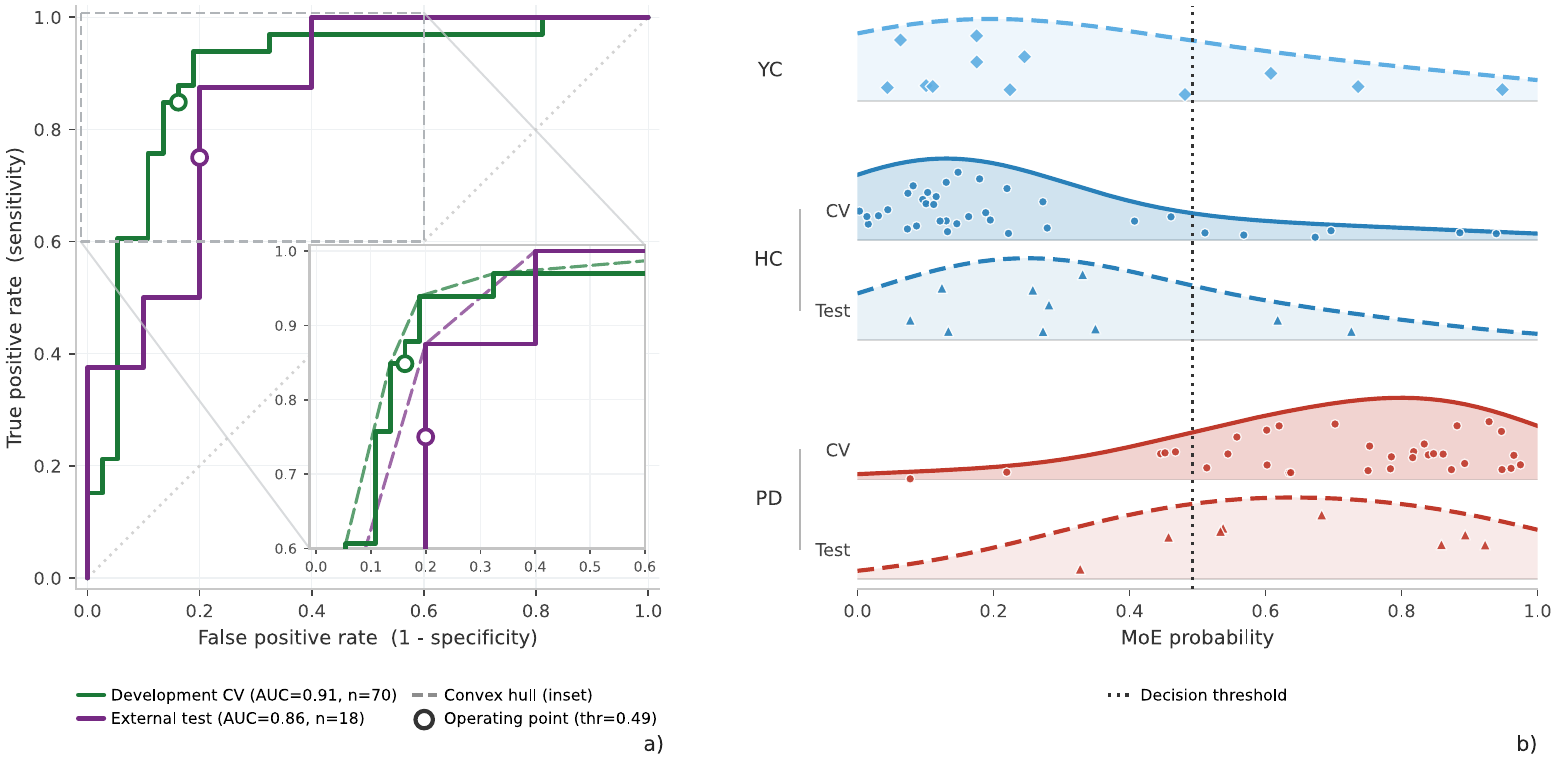}
\caption{Performance of the selected pipeline. (a) \ac{ROC} curves; the development curve uses the mean validation probability of each participant across the 20 holdouts and the inset magnifies the upper-left corner. (b) \ac{MoE} probabilities of the \ac{YC} cohort and of the \acp{HC} and participants with \ac{PD} in the development set (CV) and the external test set. The \ac{YC} cohort is compared with the external test set because neither was used to develop the pipeline. Symbols denote individual participants and curves show density estimates.}
\label{fig:selected-configuration-performance}
\end{manuscriptfigure}

The \ac{YC} cohort was not used to develop the pipeline, so its probabilities were compared with those of the external test set rather than with the development set. Nine out of 12 \acp{YC} were classified as \acp{HC}. Their median probability was \(0.20\), compared with \(0.28\) for the external \acp{HC} and \(0.61\) for the external participants with \ac{PD} (Fig.~\ref{fig:selected-configuration-performance}b).

\subsection*{Base models and features of the selected pipeline}

\paragraph{Base models.}
Fifteen out of 24 base models contributed to the combined output. The base models along the X axis accounted for \(88.58\%\) of the total contribution, with the largest contributions coming from those for \ac{SPT}5, \ac{SPT}4, \ac{SPT}9 and \ac{SPT}1 (Table~\ref{tab:expert-contribution-by-spt-and-axis}). The highest development \acp{AUC} were obtained by the base models along the X axis for \ac{SPT}4, \ac{SPT}5 and \ac{SPT}1. Base models along the X axis generally outperformed those along the Y axis, including during vertical target movement, while most base models along the Y axis performed close to chance (Supplementary Table~\ref{tab:supp-expert-auc}).

\begin{manuscripttable}
\centering
\caption{Contribution (\%) of the individual base models to the combined output, grouped by target movement. X and Y indicate the axis of each base model. Values of \(0.00\) correspond to base models without influence on the combined output.}
\label{tab:expert-contribution-by-spt-and-axis}
\uniformtablesetup

\begin{tabular*}{\textwidth}{@{\extracolsep{\fill}}ccccccccc@{}}
\toprule
\multicolumn{3}{c}{\textbf{Target movement: X}} &
\multicolumn{3}{c}{\textbf{Target movement: Y}} &
\multicolumn{3}{c}{\textbf{Target movement: 2D}} \\
\cmidrule(lr){1-3}
\cmidrule(lr){4-6}
\cmidrule(lr){7-9}

\textbf{\acs{SPT}} & \textbf{X} & \textbf{Y} &
\textbf{\acs{SPT}} & \textbf{X} & \textbf{Y} &
\textbf{\acs{SPT}} & \textbf{X} & \textbf{Y} \\
\midrule

\ac{SPT}1 & 12.13 & 0.00 &
\ac{SPT}5 & 20.87 & 9.94 &
\ac{SPT}9 & 18.27 & 1.41 \\

\ac{SPT}2 & 0.80 & 0.00 &
\ac{SPT}6 & 2.37 & 0.07 &
\ac{SPT}10 & 2.67 & 0.00 \\

\ac{SPT}3 & 6.84 & 0.00 &
\ac{SPT}7 & 2.44 & 0.00 &
\ac{SPT}11 & \(<0.01\) & 0.00 \\

\ac{SPT}4 & 18.52 & 0.00 &
\ac{SPT}8 & 2.55 & 0.00 &
\ac{SPT}12 & 1.12 & 0.00 \\

\bottomrule
\end{tabular*}
\end{manuscripttable}

\paragraph{Features.}
The eight measures with the highest contribution accounted for \(75.34\%\) of the total (Table~\ref{tab:individual-feature-contribution-selected-configuration}). Ordinal and permutation entropy features accounted for \(48.45\%\) of the total contribution.

\begin{manuscripttable}
\centering
\caption{Eight measures with the highest contribution to the combined output, reported separately for each eye. Each value combines the X and Y axes. Contr.: contribution, expressed as a percentage of the total across all features and base models; the total adds both eyes and determines the order of the rows. Direction: diagnostic group with which the model associated higher values and, in parentheses, number of the 15 active base models with that direction. \(d\): median Cohen's \(d\) across the active base models, computed as the difference between the \ac{PD} and \ac{HC} means divided by the pooled standard deviation.}
\label{tab:individual-feature-contribution-selected-configuration}
\uniformtablesetup
\begin{tabularx}{\textwidth}{@{}>{\RaggedRight\arraybackslash}p{0.17\textwidth} *{7}{>{\centering\arraybackslash}X}@{}}
\toprule
& & \multicolumn{3}{c}{\textbf{Left eye}} & \multicolumn{3}{c}{\textbf{Right eye}} \\
\cmidrule(lr){3-5}\cmidrule(l){6-8}
\textbf{Measure} &
\makecell[c]{\textbf{Total}\\\textbf{(\%)}} &
\makecell[c]{\textbf{Contr.}\\\textbf{(\%)}} & \textbf{Direction} & \textbf{\(d\)} &
\makecell[c]{\textbf{Contr.}\\\textbf{(\%)}} & \textbf{Direction} & \textbf{\(d\)} \\
\midrule

\acs{wPE} &
\(13.40\) &
\(7.55\) & \acs{PD} (\(13/15\)) & \(0.45\) &
\(5.85\) & \acs{PD} (\(12/15\)) & \(0.36\) \\

\acs{HE} &
\(10.81\) &
\(7.81\) & \acs{HC} (\(14/15\)) & \(-0.47\) &
\(3.00\) & \acs{PD} (\(10/15\)) & \(0.03\) \\

\acs{LZC} &
\(10.42\) &
\(6.41\) & \acs{HC} (\(14/15\)) & \(-0.16\) &
\(4.01\) & \acs{PD} (\(8/15\)) & \(0.15\) \\

Conditional \acs{UQPE} &
\(9.51\) &
\(4.95\) & \acs{PD} (\(9/15\)) & \(0.29\) &
\(4.56\) & \acs{PD} (\(8/15\)) & \(0.20\) \\

\acs{CD} &
\(8.71\) &
\(3.11\) & \acs{PD} (\(13/15\)) & \(0.07\) &
\(5.60\) & \acs{PD} (\(10/15\)) & \(0.07\) \\

\acs{DiffEn} &
\(8.52\) &
\(4.49\) & \acs{PD} (\(10/15\)) & \(0.15\) &
\(4.03\) & \acs{PD} (\(13/15\)) & \(0.24\) \\

\acs{FGPE} &
\(7.02\) &
\(2.46\) & \acs{HC} (\(12/15\)) & \(-0.16\) &
\(4.56\) & \acs{HC} (\(12/15\)) & \(-0.32\) \\

\acs{KFD} &
\(6.96\) &
\(2.86\) & \acs{PD} (\(9/15\)) & \(0.20\) &
\(4.10\) & \acs{PD} (\(11/15\)) & \(0.28\) \\

\bottomrule
\end{tabularx}
\end{manuscripttable}

The \ac{HE} of the left eye was the individual feature with the highest contribution. Higher values were associated with \acp{HC} in 14 out of 15 active base models and Cohen's \(d\) was negative in all 15, whereas the \ac{HE} of the right eye showed no difference between diagnostic groups. The \ac{wPE} was higher in participants with \ac{PD} in both eyes. Across all base models and features, the direction of the association agreed with the sign of Cohen's \(d\) for \(82.95\%\) of the total contribution.

\section*{Discussion and Conclusions}

On the external test set, the selected pipeline obtained an \ac{AUC} of \(0.86\). Its most consistent feature pattern was lower temporal persistence in participants with \ac{PD}, measured by the \ac{HE} of the left eye and complemented by changes in ordinal organisation captured mainly by \ac{wPE}.

\paragraph{Interpretation of the selected experimental setting.}

With \ac{DS}1, an \ac{EDel} of 25~ms and an \ac{EDim} of 3 or 4 gave delay vectors spanning 50--75~ms, whereas the other strategies spanned hundreds of milliseconds to seconds and obtained lower \ac{AUC} with the features computed from delay vectors (Supplementary Table~\ref{tab:supp-delay-strategy-selection}). Short temporal relations were therefore the most discriminative within this dataset, but this does not establish this scale as physiologically optimal. Ideally, this scale should be shorter than the oculomotor events of interest, so that each delay vector describes the dynamics within an event. Saccades typically last 30--80~ms~\cite{orbandexivry2007saccades}, so a span of 50--75~ms may combine parts of a catch-up saccade with the surrounding pursuit~\cite{DeBrouwer2002CatchUpSaccades}.

\ac{BW} removal strongly reduced global dispersion while largely preserving local variation (Supplementary Table~\ref{tab:supp-framework-sequence-properties}). The \ac{BW} arises from sources such as calibration and head movement and can mask faster fluctuations, so its removal has been reported as a necessary step for the automatic analysis of \ac{SPEM} sequences~\cite{Bejani2023BaselineWR}. The higher \ac{AUC} of \ac{EF}2 is consistent with these fluctuations carrying discriminative information, although whether they arise from corrective movements, pursuit variability or measurement noise cannot be determined here. The effect of \ac{BW} removal on Cohen's \(d\) depended on whether a measure is sensitive to amplitude (Supplementary Table~\ref{tab:supp-family-level-framework-selection}). \ac{PowEn}, which describes the amplitude distribution and was grouped with \ac{DiffEn} by \ac{HFS}, lost most of its group difference (from \(0.53\) to \(0.08\)). \ac{SampEn}, which uses a tolerance proportional to the standard deviation and is therefore insensitive to amplitude~\cite{Richman2000SampEn}, kept it (from \(0.38\) to \(0.35\)).

\ac{CERM} extraction changes the sequence in two ways. It is obtained from whitened sequences and normalised to \([0,1]\), which removes the amplitude differences between participants, and it discards the component not shared by both eyes, which contains recording artefacts and the effects of convergence deficits and eye dominance~\cite{luque2024estimation}. It also reduced the relative local variation and coincided with lower discrimination in \ac{EF}3 and \ac{EF}4. Both changes are reflected in Cohen's \(d\): the group difference of \ac{PowEn} reversed, consistent with the loss of amplitude information, and that of \ac{SampEn} also reversed, although it is insensitive to amplitude. The reversal of \ac{SampEn} is consistent with part of the more irregular local fluctuations in \ac{PD} being specific to each eye and therefore removed with the non-shared component. These associations, however, do not establish that the transformations caused the reduction in performance.

Base models were separated by axis but not by eye. The two axes differed widely in discrimination, whereas the two eyes gave similar results along the X axis, as expected from their conjugate movement during pursuit, so separate base models for each eye would have doubled the number of base models without adding discriminative information (Supplementary Table~\ref{tab:supp-channel-decomposition}). Sharing one feature subset and one \ac{SVM} configuration across base models limited the number of choices fitted to 70 participants, at the cost of not adapting the feature subset to each \ac{SPT} and axis.

\paragraph{Feature selection.}

With the complete feature set, redundancy distorted the interpretation of the model. Groups of correlated measures accumulated more contribution than their representatives because the contribution was spread over the features of several nearly identical measures. Nearly identical features were also sometimes used in opposite directions. Both effects are expected when the weight of collinear features is not identifiable~\cite{buhlmann2013correlated}. \ac{HFS} removed this redundancy without reducing development performance, so the contributions of the selected model can be interpreted feature by feature. \ac{FFS} then retained the representatives that improved classification. The selected subset outperformed random subsets of the same size, so its benefit did not arise only from using fewer measures (Table~\ref{tab:performance-summary}).

\paragraph{Interpretation of the selected measures.}

A feature is described as higher or lower in participants with \ac{PD} when its median Cohen's \(d\) is positive or negative, respectively, and a measure when this holds for the features of both eyes (Table~\ref{tab:individual-feature-contribution-selected-configuration}).

The \ac{HE} of the left eye had the highest contribution and was lower in participants with \ac{PD} consistently across the active base models, whereas the \ac{HE} of the right eye showed no difference between diagnostic groups. This difference between eyes may relate to ocular dominance, the clinically more affected side or altered binocular coordination in \ac{PD}~\cite{wu2018eye}; none of these factors was recorded or analysed. Values remained above \(0.50\) in both diagnostic groups, so the lower \ac{HE} indicates weaker temporal persistence rather than anti-persistence~\cite{Amor2016Persistence}: the gaze maintained its direction of movement for shorter periods. This may reflect shorter periods of stable pursuit, more frequent transitions between smooth tracking and correction or altered organisation of successive corrections, which is compatible with previous reports of reduced gain, altered velocity and phase and fragmented pursuit in \ac{PD}~\cite{frei2021abnormalities,Popovic2026SmoothPursuitMemorySaccades}. Evidence of altered basal ganglia output and impaired extraretinal pursuit in \ac{PD} is consistent with less stable automatic or predictive control, although the present data do not identify its neural origin~\cite{Obeso2008BasalGanglia,Fukushima2015ImpairedPursuit}.

\ac{wPE}, \ac{DiffEn} and \ac{KFD} were higher in participants with \ac{PD} in both eyes. Higher \ac{wPE} indicates less predictable local fluctuations~\cite{Fadlallah2013wPE}, higher \ac{DiffEn} a wider distribution of gaze positions~\cite{Shannon1948} and higher \ac{KFD} a more irregular waveform~\cite{katz1988fractals}. In terms of tracking, the three measures describe gaze that fluctuates more irregularly. Conditional \ac{UQPE} was also higher in both eyes, but its model direction was inconsistent across base models. Together with the lower \ac{HE} of the left eye, these measures describe gaze position sequences in \ac{PD} that are less persistent and more irregular at short time scales, a profile consistent with less stable pursuit interrupted by frequent small corrections. The measures, however, do not identify the events involved.

\ac{FGPE} was lower in participants with \ac{PD} in both eyes, in the opposite direction to \ac{wPE}, so ordinal and permutation entropy measures cannot be summarised as higher or lower complexity in \ac{PD}. \ac{LZC} showed small effects of opposite sign in the two eyes and \ac{CD} showed no difference between diagnostic groups, although both contributed to the model. Their contribution therefore arose in combination with the other features rather than from a group difference of their own.

\paragraph{Distribution of information across axes and \acp{SPT}.}

The X axis predominated in the discrimination of individual base models and in their contribution, including during vertical trajectories (Table~\ref{tab:expert-contribution-by-spt-and-axis}; Supplementary Table~\ref{tab:supp-expert-auc}). During vertical pursuit, it may capture transverse deviations, corrections or saccadic intrusions around the intended trajectory. Normal directional differences in pursuit, target dynamics, calibration, head position and measurement precision remain alternative explanations~\cite{Mcilreavy2019TwoDimensional,Danion2021DirectionalAsymmetry}. The X axis was therefore the most informative source under this protocol, which does not imply a preferential impairment of horizontal pursuit. Because the 12 \acp{SPT} were administered in a fixed order, differences between \acp{SPT} cannot be attributed exclusively to their trajectory, frequency or amplitude; adaptation, attention, fatigue or learning may have contributed.

\paragraph{External evaluation and clinical scope.}

The external \ac{AUC} had a wide confidence interval because the test set contained 18 participants. Sensitivity and specificity were similar, so the classification was balanced between diagnostic groups, as expected from the \ac{EER} threshold. This supports generalisation to unseen participants under comparable acquisition conditions, but does not establish transportability across centres, devices, operators or implementations. Screening, in which false negatives have a greater clinical cost, would require a prospectively defined threshold that favours sensitivity.

The distributions of the \acp{YC} and the external \acp{HC} overlapped and 9 out of 12 \acp{YC} were classified as \acp{HC} (Fig.~\ref{fig:selected-configuration-performance}b). This provides no indication that predictions for healthy participants depended on age, although age and cognition were not modelled directly. This comparison only describes predictions for younger healthy participants; \ac{PD}--\ac{HC} discrimination was assessed on the external test set.

Other automatic studies used different eye movement tasks, technologies, outcomes, validation designs and units of analysis, which precludes direct numerical comparison~\cite{bejani2022detection,brien2023classification,Jiang2025Diagnosis,Zhang2026VR,Tan2026DPMDLA}. The present study combines participant classification with an analysis of the temporal and spatial properties that support the predictions. Because eye movement abnormalities also occur in other neurological and Parkinsonian disorders, \ac{PD}--\ac{HC} discrimination does not establish diagnostic specificity~\cite{Antoniades2024EyeMovements}.

\paragraph{Limitations and future work.}

The same 70 participants were used to select the feature subset, the \ac{SVM} configuration, the combiner and the threshold, so the development estimates may overestimate performance on new participants. The external test set limits this risk, but its size prevents a precise estimate of the difference. The higher proportion of exclusions in the \ac{PD} group limits applicability to participants with \ac{PD} who can complete the protocol and provide technically valid recordings. All recordings were acquired with the same protocol, equipment and clinical team, so performance with other recording systems, protocols or clinical settings remains untested. Finally, the values of complexity and nonlinear dynamics measures depend on the length of the sequence, the parameters of each measure and the noise. A positive \ac{LLE} in a finite, noisy and nonstationary recording, for example, does not demonstrate deterministic chaos~\cite{Melnyk2024Entropy}. The selected measures should therefore be interpreted as empirical descriptions of the gaze position sequences rather than as direct measurements of oculomotor control.

Larger studies with repeated measurements should assess associations with age, severity, medication, cognition and the clinically more affected side, together with feature and probability stability~\cite{Dalbro2025Repeatability}. Event annotation and comparison with the target position sequence are needed to locate the observed patterns within smooth tracking, corrections or their transitions. Independent recruitment, other recording systems and neurological comparison groups are required to establish transportability and specificity.

\paragraph{Conclusions.}

The results support the feasibility of using complexity and nonlinear dynamics measures of gaze position sequences to distinguish participants with early \ac{PD} from \acp{HC}. Within this dataset, the most informative representation combined a short reconstruction scale with \ac{BW} removal and the separate gaze position sequences of the two eyes. The most important features described gaze position sequences in \ac{PD} that were less persistent and fluctuated more irregularly at short time scales, a profile consistent with less stable pursuit. Evaluation on the external test set supported generalisation under comparable acquisition conditions, but its small size limits the precision of the estimate. The proposed pipeline should therefore be considered a candidate tool for screening and biomarker discovery rather than a replacement for clinical diagnosis.

\FloatBarrier

\paragraph{Acknowledgements.}
The authors thank the participants; the Neurology Service of Hospital General Universitario Gregorio Marañón for its facilities; the Madrid ELLIS unit for indirect support; and Universidad Politécnica de Madrid for Magerit Supercomputer resources. Funding was provided by the Ministry of Economy and Competitiveness of Spain (DPI2017-83405-R1 and PID2021-128469OB-I00) and Comunidad de Madrid. Universidad Politécnica de Madrid supports Julián D. Arias-Londoño through the María Zambrano grant UP2021-035, funded by the European Union--NextGenerationEU.

\paragraph{CRediT Author Statement.}
\begin{sloppypar}
Conceptualisation and Methodology: J.D.A., J.I.G.; Software: J.C.P., M.F.A., J.D.A.; Validation: J.C.P., M.F.A., J.D.A., J.I.G.; Investigation (participant recruitment and data acquisition): E.L.B., F.G.P.; Resources: F.G.P., J.I.G.; Writing---original draft: J.C.P., J.I.G.; Writing---review and editing: J.D.A., J.I.G.; Supervision: F.G.P., J.D.A., J.I.G.; Project administration and Funding acquisition: J.I.G. All authors read and approved the final manuscript.
\end{sloppypar}

\paragraph{Declaration of Competing Interests.}
The authors declare no competing interests.

\prepareappendix
\appendix

\setcounter{figure}{0}
\renewcommand{\thefigure}{\thesection.\arabic{figure}}
\setcounter{table}{0}
\renewcommand{\thetable}{\thesection.\arabic{table}}

\section{Embedding delay and dimension estimation}
\label{app:embedding-parameters}

This appendix reports the \ac{EDel}, \ac{EDim} and temporal span obtained with each \ac{DS} and explains why \ac{DS}1 was used in the final pipeline.

\subsection{Illustration of the delay strategies}
\label{app:delay-strategy-illustration}

Figure~\ref{fig:appendix-delay-strategies} shows the four \acp{DS} for one \ac{HC} and one participant with \ac{PD} in \ac{SPT}11. \ac{DS}1 used 25~ms in both participants, whereas \ac{DS}2--\ac{DS}4 gave delays between 173 and 789~ms. The data-driven strategies therefore placed consecutive coordinates of each delay vector hundreds of milliseconds apart, whereas \ac{DS}1 kept them within a few tens of milliseconds.

\begin{appendixfigure}
\centering
\includegraphics[
width=0.90\linewidth
]{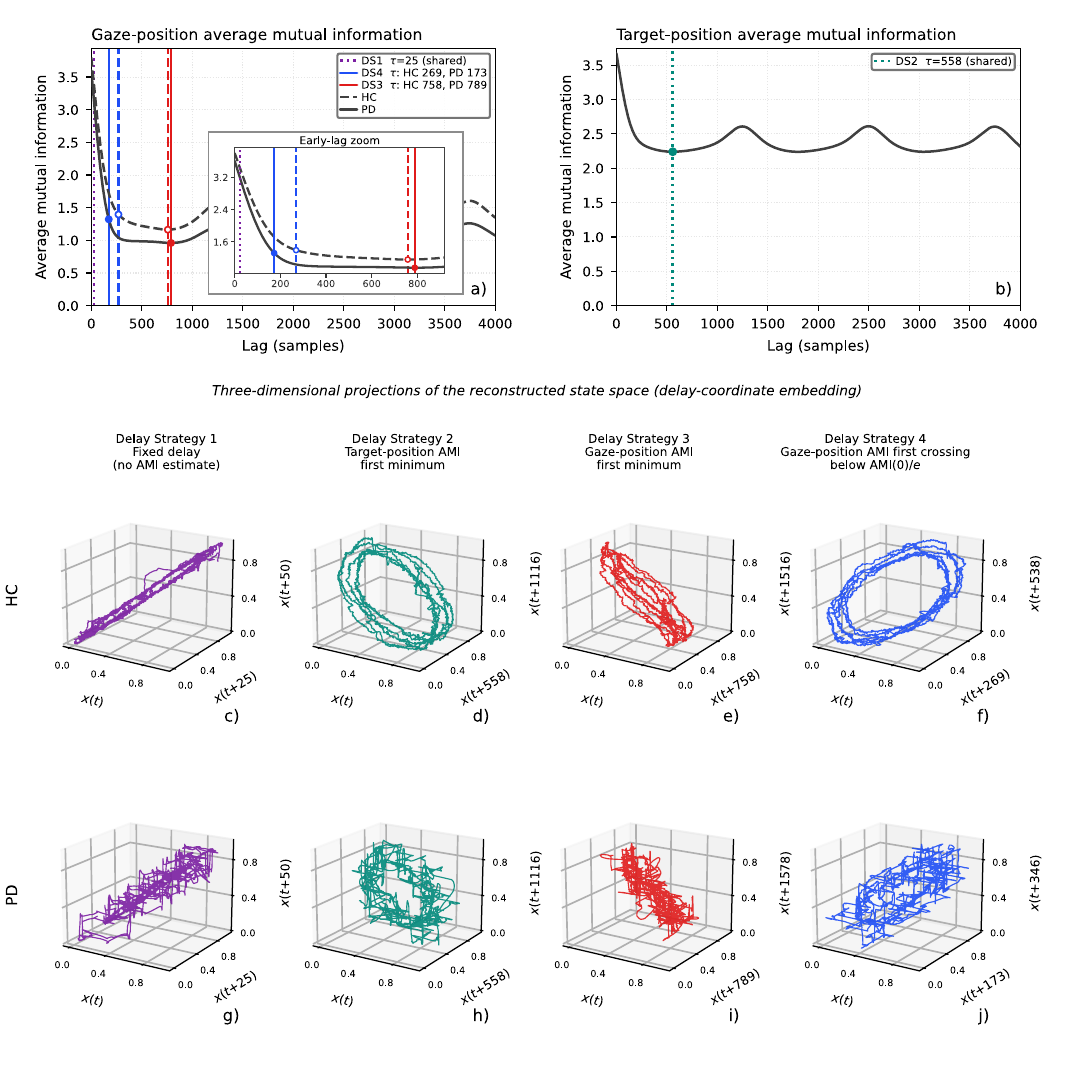}
\caption{Four \acp{DS} for participants HG060 (\ac{HC}) and HF014 (\ac{PD}) in \ac{SPT}11, left eye, X axis, before \ac{BW} removal. (a) \ac{AMI} curves of the gaze position sequences for \ac{HC} (dashed) and \ac{PD} (solid); coloured lines denote the selected delays. (b) \ac{AMI} curve of the target position sequence; the coloured line denotes the shared \ac{DS}2 delay. (c--f) Three-dimensional phase space reconstructions for \ac{DS}1--\ac{DS}4 in HG060 (\ac{HC}). (g--j) Corresponding phase space reconstructions for HF014 (\ac{PD}).}
\label{fig:appendix-delay-strategies}
\end{appendixfigure}

\subsection{Embedding delay of the data-driven strategies}
\label{app:ds2-ds4-values}

In \ac{DS}2, the first \ac{AMI} minimum of the target position sequence gave an \ac{EDel} between 558~ms and 1.02~s, which corresponded to \(20\%\)--\(22\%\) of the target period in every \ac{SPT} and axis with target movement. \ac{DS}2 therefore tied the delay to the target period rather than to the gaze position sequence.

\ac{DS}3 and \ac{DS}4 gave one \ac{EDel} for each participant and sequence (Fig.~\ref{fig:appendix-ds3-ds4-distributions}). Median values ranged from 448 to 775~ms for \ac{DS}3 and from 115 to 472~ms for \ac{DS}4. Both strategies therefore produced delays about 5 to 30 times longer than \ac{DS}1 that differed between participants, so the features computed from delay vectors described a different time scale for each participant.

\begin{appendixlatefigure}
    \centering
    \includegraphics[width=0.92\linewidth]{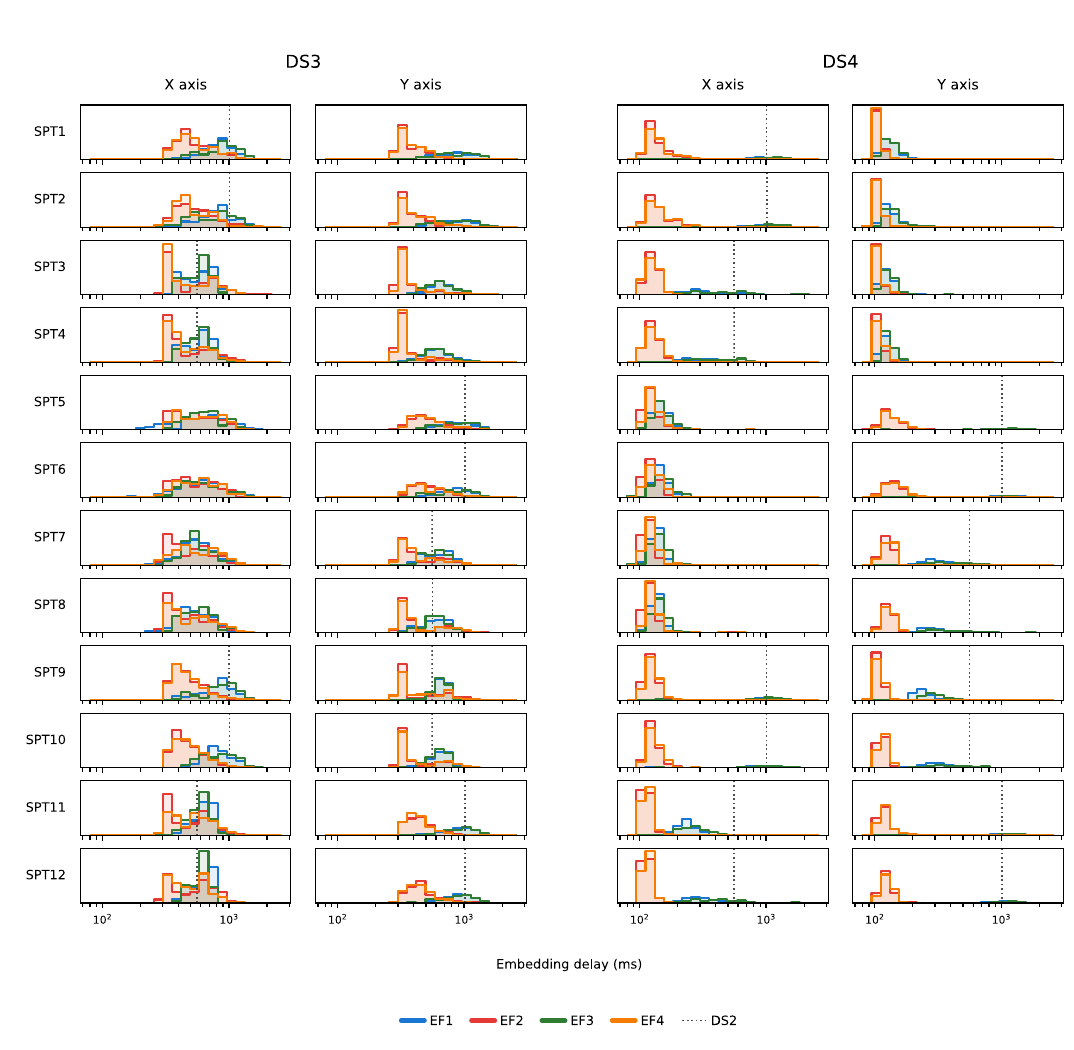}
    \caption{\ac{EDel} distributions for \ac{DS}3 (first two columns) and \ac{DS}4 (last two) by \ac{SPT} and axis. Colours denote \ac{EF}1--\ac{EF}4; dotted lines show \ac{DS}2. The horizontal axis is logarithmic and expressed in milliseconds.}
    \label{fig:appendix-ds3-ds4-distributions}
\end{appendixlatefigure}

\subsection{Embedding dimension and temporal span}
\label{app:embedding-dimension-temporal-span}

\ac{FNN} estimated an \ac{EDim} between 2 and 4 with \ac{DS}1 and between 2 and 9 with the other strategies (Fig.~\ref{fig:appendix-embedding-dimension-temporal-span}). Combined with each \ac{EDel} through Eq.~\eqref{eq:temporal-span}, this gave median temporal spans of 50--75~ms for \ac{DS}1, compared with 2.02--2.24~s for \ac{DS}2, 2.03--3.20~s for \ac{DS}3 and 468~ms--1.60~s for \ac{DS}4.

\begin{appendixassetfigure}
    \centering
    \includegraphics[
        width=0.94\linewidth
    ]{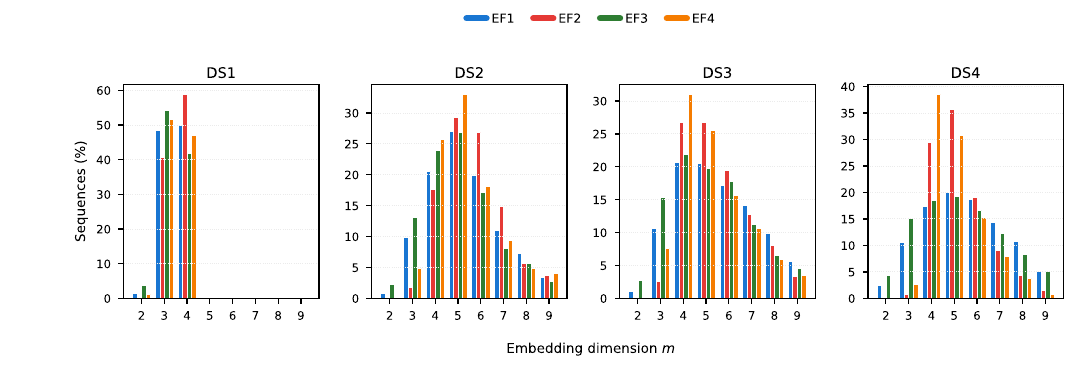}
    \caption{\ac{EDim} distributions for \ac{DS}1--\ac{DS}4. Colours denote \ac{EF}1--\ac{EF}4, and bars show the percentage of sequences for each embedding dimension.}
    \label{fig:appendix-embedding-dimension-temporal-span}
\end{appendixassetfigure}

\ac{DS}1 was therefore the only strategy whose delay vectors covered tens of milliseconds. The other strategies covered a substantial fraction of the target period, a scale at which the delay vectors are expected to reflect mainly the slow trajectory imposed by the target. Together with its higher \ac{AUC} in the preliminary analysis (Table~\ref{tab:experimental-setting-summary}), this motivated the use of \ac{DS}1 in the final pipeline, with an \ac{EDel} of 25~samples and the \ac{EDim} estimated for each sequence.

\FloatBarrier

\startreferencecolumns
\bibliographystyle{unsrt}
\bibliography{sample}
\stopreferencecolumns

\clearpage
\acresetall
\section*{Supplementary results}

\setcounter{table}{0}
\renewcommand{\thetable}{S\arabic{table}}
\setcounter{figure}{0}
\renewcommand{\thefigure}{S\arabic{figure}}

\FloatBarrier
\subsection*{S1. Preliminary selection of the experimental setting}

\begin{table}[!htbp]
\centering
\caption{Development set \ac{AUC} for \ac{PD} classification across the \acp{DS} using the gaze position sequences along the X axis. Values are reported as the mean $\pm$ standard deviation across the four \acp{EF}.}
\label{tab:supp-delay-strategy-selection}
\uniformtablesetup
\begin{tabular*}{\textwidth}{@{\extracolsep{\fill}}ccccc@{}}
\toprule
\textbf{\acs{DS}} & \makecell[c]{\textbf{Approximate}\\\textbf{temporal span}} & \makecell[c]{\textbf{All}\\\textbf{features}} & \makecell[c]{\textbf{Computed from}\\\textbf{delay vectors}} & \makecell[c]{\textbf{Computed directly}\\\textbf{from the sequence}} \\
\midrule
\textbf{\ac{DS}1} & \textbf{50--75~ms} & \(\mathbf{0.68 \pm 0.07}\) & \(\mathbf{0.65 \pm 0.14}\) & \(0.64 \pm 0.10\) \\
\ac{DS}2 & approximately 2.20~s & \(0.66 \pm 0.07\) & \(0.62 \pm 0.04\) & \(0.64 \pm 0.10\) \\
\ac{DS}3 & approximately 3.20~s & \(0.67 \pm 0.06\) & \(0.63 \pm 0.04\) & \(0.64 \pm 0.10\) \\
\ac{DS}4 & approximately 1.50~s & \(0.61 \pm 0.08\) & \(0.56 \pm 0.11\) & \(0.65 \pm 0.10\) \\
\bottomrule
\end{tabular*}
\end{table}

\begin{table}[!htbp]
\centering
\caption{Development set \ac{AUC} for \ac{PD} classification according to the analysed axis and the eye from which the features were extracted at \ac{DS}1. Values are reported as the mean $\pm$ standard deviation across the four \acp{EF}. Separate eyes along the Y axis are not shown; their \ac{AUC} was close to chance (approximately \(0.50\)).}
\label{tab:supp-channel-decomposition}
\uniformtablesetup
\begin{tabularx}{\textwidth}{@{}>{\RaggedRight\arraybackslash}Xcc@{}}
\toprule
\textbf{Data included at \ac{DS}1} &
\textbf{All features} &
\textbf{Features computed from delay vectors} \\
\midrule

Both eyes, X and Y axes pooled &
\(0.66 \pm 0.07\) &
\(0.64 \pm 0.12\) \\

Both eyes, X axis &
\(\mathbf{0.68 \pm 0.07}\) &
\(\mathbf{0.65 \pm 0.14}\) \\

Both eyes, Y axis &
\(0.51 \pm 0.06\) &
\(0.55 \pm 0.05\) \\

Left eye, X axis &
\(0.68 \pm 0.07\) &
\(0.64 \pm 0.14\) \\

Right eye, X axis &
\(0.70 \pm 0.08\) &
\(0.66 \pm 0.14\) \\

\bottomrule
\end{tabularx}

\end{table}

\begin{table}[!htbp]
\centering
\caption{Changes in the gaze position sequences of the development set produced by \ac{BW} removal and \ac{CERM} extraction. Relative local variation is the standard deviation of the first differences divided by the standard deviation of the sequence. Values are median paired ratios \(R=M_{\mathrm{after}}/M_{\mathrm{before}}\) with \(95\%\) confidence intervals; \(R<1\) indicates a reduction.}
\label{tab:supp-framework-sequence-properties}
\uniformtablesetup
\begin{tabularx}{\textwidth}{@{}>{\RaggedRight\arraybackslash}X
>{\centering\arraybackslash}m{0.17\textwidth}
>{\centering\arraybackslash}m{0.17\textwidth}
>{\centering\arraybackslash}m{0.17\textwidth}@{}}
\toprule
\textbf{Statistic} &
\makecell[c]{\textbf{\acs{BW} removal}\\\textbf{\acs{EF}2/\acs{EF}1}} &
\makecell[c]{\textbf{\acs{CERM} extraction}\\\textbf{\acs{EF}3/\acs{EF}1}} &
\makecell[c]{\textbf{\acs{CERM} extraction}\\\textbf{\acs{EF}4/\acs{EF}2}} \\
\midrule
Standard deviation of the sequence & \(0.17\;[0.15,\,0.19]\) & --- & --- \\
Standard deviation of the first differences & \(0.97\;[0.97,\,0.98]\) & --- & --- \\
Relative local variation, complete sequence & --- & \(0.93\;[0.92,\,0.93]\) & \(0.95\;[0.94,\,0.95]\) \\
Relative local variation, 50~ms windows & --- & \(0.81\;[0.81,\,0.81]\) & \(0.89\;[0.89,\,0.90]\) \\
Windows with \(R<1\) (\%) & --- & \(80.88\;[79.76,\,81.15]\) & \(70.70\;[69.06,\,71.87]\) \\
\bottomrule
\end{tabularx}
\end{table}

\begin{table}[!htbp]
\centering
\caption{Representative effect sizes by family at \ac{DS}1 in the development set. Cohen's \(d\) was calculated as the difference between the \ac{PD} and \ac{HC} means divided by the pooled standard deviation. Positive values therefore indicate higher feature values in participants with \ac{PD}, whereas negative values indicate higher values in \acp{HC}. Bold rows identify the measures whose direction reversed after \ac{CERM} extraction both with and without \ac{BW} removal.}
\label{tab:supp-family-level-framework-selection}
\uniformtablesetup
\begin{tabularx}{\textwidth}{@{}>{\RaggedRight\arraybackslash}Xcccc@{}}
\toprule
\textbf{Family / representative measure} &
\textbf{\ac{EF}1} &
\textbf{\ac{EF}2} &
\textbf{\ac{EF}3} &
\textbf{\ac{EF}4} \\
\midrule

Memory and scaling: \acs{DFA} &
-0.39 & -0.19 & -0.26 & -0.24 \\

Memory and scaling: \acs{HE} &
-0.16 & -0.29 & -0.11 & -0.22 \\

Symbolic complexity: \acs{LZC} &
+0.32 & -0.08 & +0.16 & -0.06 \\

Geometric and fractal: \acs{PFD} &
+0.23 & +0.18 & +0.25 & +0.16 \\

Phase space dynamics: \acs{LLE} &
+0.40 & +0.35 & +0.30 & +0.33 \\

Phase space dynamics: \acs{CD} &
+0.06 & +0.19 & -0.30 & +0.08 \\

\textbf{Informational entropy: \acs{PowEn}} &
\textbf{+0.53} &
\textbf{+0.08} &
\textbf{-0.30} &
\textbf{-0.21} \\

\textbf{Temporal entropy: \acs{SampEn}} &
\textbf{+0.38} &
\textbf{+0.35} &
\textbf{-0.18} &
\textbf{-0.25} \\

Ordinal and permutation entropy: \acs{wPE} &
+0.78 & +0.65 & +0.53 & +0.61 \\

\bottomrule
\end{tabularx}

\end{table}

\FloatBarrier
\subsection*{S2. Discriminative ability of the individual base models}

\begin{table}[!htbp]
\centering
\caption{\ac{AUC} of the base models by \ac{SPT}, target movement and axis. Cells report the mean (standard deviation) across the 20 holdouts, followed by the \ac{AUC} on the external test set. Bold values identify the three highest development values.}
\label{tab:supp-expert-auc}
\uniformtablesetup

\begin{tabularx}{\textwidth}{@{}
>{\RaggedRight\arraybackslash}p{0.13\textwidth}
>{\centering\arraybackslash}X
>{\centering\arraybackslash}X
>{\centering\arraybackslash}X
>{\centering\arraybackslash}X
@{}}
\toprule
\textbf{Target: X} &
\textbf{\acs{SPT}1} &
\textbf{\acs{SPT}2} &
\textbf{\acs{SPT}3} &
\textbf{\acs{SPT}4} \\
\cmidrule(lr){1-5}
X axis &
\(\mathbf{0.77\,(0.06)}\,/\,0.48\) &
\(0.70\,(0.11)\,/\,0.80\) &
\(0.66\,(0.15)\,/\,0.80\) &
\(\mathbf{0.82\,(0.08)}\,/\,0.70\) \\
Y axis &
\(0.47\,(0.08)\,/\,0.50\) &
\(0.51\,(0.03)\,/\,0.50\) &
\(0.47\,(0.07)\,/\,0.50\) &
\(0.53\,(0.10)\,/\,0.50\) \\

\addlinespace[3pt]
\textbf{Target: Y} &
\textbf{\acs{SPT}5} &
\textbf{\acs{SPT}6} &
\textbf{\acs{SPT}7} &
\textbf{\acs{SPT}8} \\
\cmidrule(lr){1-5}
X axis &
\(\mathbf{0.80\,(0.15)}\,/\,0.75\) &
\(0.52\,(0.15)\,/\,0.90\) &
\(0.72\,(0.15)\,/\,0.74\) &
\(0.67\,(0.11)\,/\,0.79\) \\
Y axis &
\(0.67\,(0.15)\,/\,0.64\) &
\(0.53\,(0.09)\,/\,0.40\) &
\(0.47\,(0.06)\,/\,0.50\) &
\(0.47\,(0.07)\,/\,0.50\) \\

\addlinespace[3pt]
\textbf{Target: 2D} &
\textbf{\acs{SPT}9} &
\textbf{\acs{SPT}10} &
\textbf{\acs{SPT}11} &
\textbf{\acs{SPT}12} \\
\cmidrule(lr){1-5}
X axis &
\(0.74\,(0.10)\,/\,0.64\) &
\(0.58\,(0.09)\,/\,0.60\) &
\(0.62\,(0.06)\,/\,0.71\) &
\(0.66\,(0.17)\,/\,0.80\) \\
Y axis &
\(0.55\,(0.09)\,/\,0.57\) &
\(0.42\,(0.10)\,/\,0.50\) &
\(0.48\,(0.05)\,/\,0.50\) &
\(0.49\,(0.07)\,/\,0.50\) \\

\bottomrule
\end{tabularx}
\end{table}

\FloatBarrier

\end{document}